\pdfoutput=1
\documentclass[twocolumn]{svjour3}
\smartqed
\usepackage{graphicx}
\usepackage{amsmath}
\usepackage{amssymb}
\usepackage{makecell}
\usepackage{mathptmx}
\usepackage[misc,geometry]{ifsym}

\DeclareMathOperator*{\argmin}{arg\,min}
\usepackage{hyperref}
\hypersetup{
    colorlinks=true,
    linkcolor=blue,
    filecolor=magenta,      
    urlcolor=cyan,
}

\begin{document}
\sloppy
\title{Arhuaco: Deep Learning and Isolation Based Security for
Distributed High-Throughput Computing}

\author{
    A. Gomez Ramirez \and
    C. Lara          \and 
    L. Betev         \and \\
    D. Bilanovic     \and
    U. Kebschull     \and
    for the ALICE Collaboration
}

\institute{A. Gomez Ramirez (\Letter) \and
           C. Lara          \and
           D. Bilanovic     \and
           U. Kebschull     \at
           IRI group, Goethe-University Frankfurt \\
           \email{andres.gomez@cern.ch}
           \and
           L. Betev \at
              CERN, Geneva, Switzerland
}

\date{Received: date / Accepted: date}

\maketitle
\begin{abstract}
Grid computing systems require innovative methods and tools to identify
cybersecurity incidents and perform autonomous actions i.e. without 
administrator intervention. They also require methods to isolate and trace job
payload activity in order to protect users and find evidence of malicious
behavior. We introduce an integrated approach of security monitoring via
Security by Isolation with Linux Containers and Deep Learning methods for the
analysis of real time data in Grid jobs running inside virtualized
High-Throughput Computing infrastructure in order to detect and prevent
intrusions. A dataset for malware detection in Grid computing is described. We
show in addition the utilization of generative methods with Recurrent Neural
Networks to improve the collected dataset. We present Arhuaco, a prototype
implementation of the proposed methods. We empirically study the performance of
our technique. The results show that Arhuaco outperforms other methods used in
Intrusion Detection Systems for Grid Computing. The study is carried out in the
ALICE Collaboration Grid, part of the Worldwide LHC Computing Grid.
\keywords{Grid Computing Security \and
          Intrusion Detection and Prevention \and
          Deep Learning \and
          WLCG \and
          Isolation \and
          Malware Detection
}
\end{abstract}

\section{Introduction}
\label{intro}
The Worldwide LHC Computing Grid (WLCG) is one the most remarkable examples of
High-Troughput Computing (HTC) distributed infrastructure for scientific
applications. The WLCG is the global Grid that analyzes data from the Large
Hadron Collider (LHC) at CERN, with 170 sites in 40 countries. Due to their
size, complexity, reputation and required access by Internet these systems are
continuously exposed to attackers. Authenticated users have the freedom to
execute arbitrary code and to transfer arbitrary data that is required for
their experiments. External or insider attackers may take advantage of the Grid
functionality to carry out unauthorized activities such as running malware or
mining cryptocoins. The Grid is a heterogeneous and dynamic environment where
it is difficult to adapt traditional rule based Intrusion Detection Systems
(IDS).

\par
\begin{figure}[h]
\begin{center}
\begin{minipage}{86mm}
\includegraphics[width=85mm,scale=1.0]{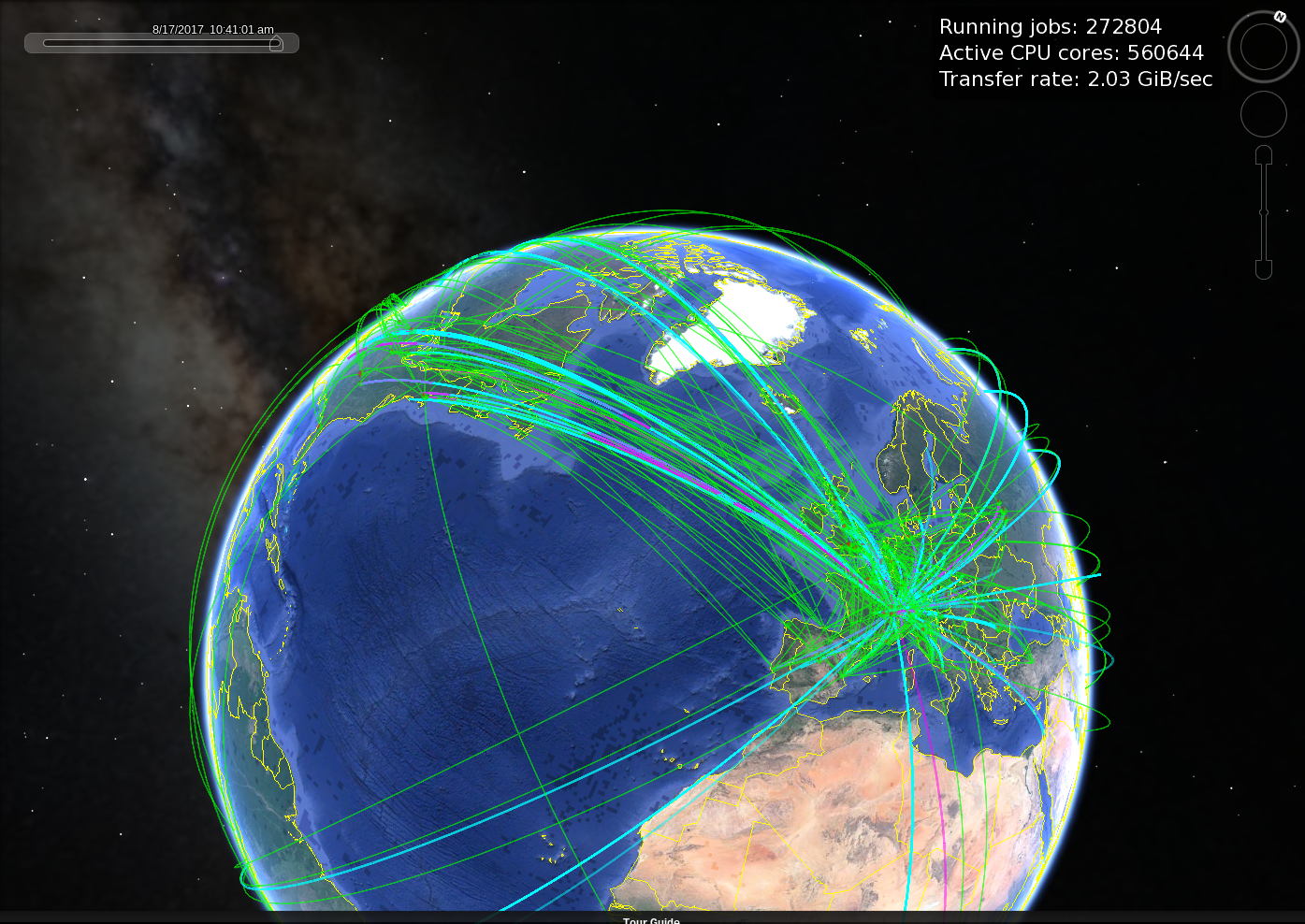}
\end{minipage}\hspace{2pc}%
\end{center}
\caption{Real time grid data flow around the world}
\label{fig:grid}
\end{figure}

The distributed usage of High-throughput Computing (HTC) farms for data
processing tasks - known as Grid Computing - has been very successful in High
Energy Physics (HEP), weather forecasting, brain research and astronomy
research, just to mention a few examples. Scientists can submit jobs
composed of custom code and experimental data. The computing Grid has been
envisioned as an analogy for the electrical Grid, for computing resources on
demand. The Worldwide LHC Computing Grid (WLCG) enables the scientists to
analyze massive amounts of physics data and it allowed for the experimental
validation of the existence of the Higgs boson \cite{grid_higss}. The WLCG
integrates computer centers worldwide that provide computing and storage
resource into a single infrastructure accessible by all Large Hadron Collider
(LHC) physicists. Currently, it combines the power of nearly 170 sites in 40
countries, connected with 10-100 Gb links, with more than 600.000 processing
cores and 700 PB of storage capacity. It is capable of processing more than 2
million jobs per day. The ALICE (A Large Ion Collider Experiment) Collaboration
has built a dedicated detector to exploit the unique physics potential of
nucleus-nucleus collisions at LHC energies. Its aim is to study the physics of
strongly interacting matter at the highest energy densities reached so far in
the laboratory \cite{Alice_experiment}. As a member of the WLCG, the ALICE
experiment has developed the ALICE production environment (AliEn)
\cite{bagnasco_alien:_2008}, implementing many components of the Grid computing
technologies that are needed to store, process and analyze the collected data.
Through AliEn, the computing centers that provide CPU and storage resources for
ALICE can be seen and used as a single entity - any available node executes
jobs and file access is transparent to the user, wherever in the world a file
might be located. Figure \ref{fig:grid} represents the flow of data in the
WLCG.

\par
Authorized users of Grid computing systems have the freedom to carry out
research on experimental data. They are able to execute arbitrary code and
transfer any required data. This means that potential insider attackers have
the same capabilities. The focus of this study is on the security issues
related to the Grid job execution environment inside site worker nodes.
Frequently in HEP Grids \cite{Grid_HEP}, the jobs running in the worker nodes
have access rights that are beyond of what is actually required, restricted
only by one or several local Linux accounts. When multiple jobs are executed
with the same account, an attacker with control over one job could tamper
another user jobs, blaming the owner of any malicious activity. These processes
could also have access to sensitive server data and restricted networks.
Therefore, Security by Isolation (SbI) mechanisms for processes and networks
are important requirements for Grid computing.

\par
An insider attacker may misuse the power of the Grid for activities not related
to physics data analysis. Complex HTC infrastructure such as WLCG are attractive
targets for external attackers as well. An attacker might take advantage of the
Grid functionality to tamper with user jobs, escalate privileges, access
sensitive server configuration data, setup a Denial of Service (DoS) attack
or mine cryptocoins, to name a few of the possibilities. This could be
accomplished by exploiting unknown or unfixed software or hardware
vulnerabilities, listen to user network traffic to gather sensitive clear text
information or by guessing weak user credentials among other possibilities.
Millions of jobs might be running in Grids like the WLGC every day. The user's
ability to run arbitrary code and the lack of proper isolation make detecting
intrusions a more challenging task than in other systems. Traditional Intrusion
Detection Systems (IDS) require rules written by security professionals. Those
rules need to be updated constantly which requires an important amount of
effort. Rule based IDS are difficult to adapt to very dynamic environments such
as the Grid \cite{gomez1}. Machine Learning algorithms can help to automate the
way IDS are built and updated, by using security monitoring data collected from
the protected systems \cite{tsai_intrusion_2009}. In the Grid, the amount of
this monitoring data is huge due to the number of running jobs. Deep Learning
methods may help to analyze this data to improve the IDS detection accuracy.
\par
We introduce Arhuaco, a framework that adopts Linux Container (LC) technology
to provide SbI and security monitoring by applying Deep Learning for detection
and prevention of abnormal activities based on multiple sources of monitoring
data such as network connections and system calls. Arhuaco gives researches the
ability to generate complementary training data by a Recurrent Neural Network.
This can be used to improve the detection performance and adapt the detector to
new environments. A dataset for Machine Learning (ML) training of malware
detection on Linux based Grid computing is described. Some of the most popular
Grid systems are just starting to explore the usage of SbI and IDS for security
monitoring \cite{wlcg}. As we describe in the related work section there are no
studies that leverage the capabilities of Linux Container isolation and
monitoring in combination with Deep Learning and data generation for Intrusion
Detection. We also show in the next section that there is no tool implementing
the mentioned techniques for Grid job payload monitoring in order to detect
intrusions. We describe the design, implementation decisions and tests of our
proposed methods in the ALICE Collaboration Grid, part of the WLCG. We
demonstrate that the selected algorithms and techniques outperform other methods
used on IDS for Grid Computing.

\par
This document is organized as follows. Section \ref{related} presents the state
of the art on ML and isolation based security methods applied to distributed
environments especially in Grid computing. Section \ref{background} provides
background information on the SbI approach and an overview of the classification
task in Machine Learning and generative models. Section \ref{design} and
\ref{implementation} describe the Arhuaco design and implementation based on
our ideas. Section \ref{results} shows the results obtained from testing our
approach. Finally section \ref{discusion} and \ref{conclusions} summarize our
findings and indicate directions of our future research.

\section{Related work}
\label{related}

\subsection{Security by Isolation in Grid Computing}
Virtual Machines (VM) have been suggested many times to solve the isolation
problems in Grid computing \cite{VMS}. VMs are emulated machines with their own
kernel while Linux Containers (LC) can share a single kernel. Several 
researchers - \cite{gomez1} and \cite{gomez2} - have proposed the usage of LCs
to provide a level of isolation between the Grid jobs and the underlying system
and network. Saving system resources in High-throughput Computing (HTC)
applications is critical, and LC help to reduce the overall performance impact. 
\cite{Xavier:2013:PEC:2497369.2497577} presents a comparison of the performance
of several virtualization technologies including VM, and shows that container
based systems have a near-native performance of CPU, memory, disk and network.
\cite{silberschatz_operating_2013} analyzes LCs and VMs and finds similar
results in terms of performance and scalability.
\cite{berzano_ground-up_nodate-2} shows a success real experience for LCs
providing isolation in a Grid site at the ALICE High Level Trigger (HLT). Our
study further extends this direction. In particular, we are interested in how
this isolation mechanism can be integrated with a security monitoring system,
that provides methods to enforce Intrusion Prevention and Detection in Grid
computing.

\subsection{Intrusion Detection}
In \cite{liao_intrusion_2013} an extensive review of Intrusion Detection
Systems is presented. Intrusion is defined as the attempt to compromise 
confidentiality, integrity and availability and Intrusion Detection as the
process of monitoring the events occurring in a computer system or network, and
analyzing them for signs of intrusions. The cited study presents several open
source technologies as the most used solutions for IDS such as SNORT
\cite{snort} and OSSEC \cite{ossec}. False positive and false negative are two
very common metrics to assess the degree of accuracy. Relevant features can be
sets of audit trails (e.g. system logs, system commands) on a host, network
packets or connections, wireless network traffic and application logs.

\par
Machine Learning has been proposed in many studies to improve IDS.
\cite{tsai_intrusion_2009} summarizes the state of the art on ML techniques
applied to Intrusion Detection and prevention. It states that the most commonly
used techniques in the topic have been K-nearest neighbor (K-NN), Support
Vector Machines (SVM), Artificial Neural Networks, self-organizing maps,
decision trees, Na\"ive Bayes networks, genetic algorithms and fuzzy logic for
one single classifier approaches. On the other hand for hybrid classifiers, using
several classifiers, neuro-fuzzy techniques, clustering-based approaches have
been used especially for parameter tunning and classification. Single
classifiers with K-NN and SVM are very popular, mainly the second one. For
hybrid approaches an integrated framework, where a method is used for feature
selection while another method is used for classification is common. KDD99
\cite{kdd99} is presented as the standard database for testing ML based IDS.
\cite{wu_use_2010} shows an overview on the usage of computational intelligence
research on IDS. According to the review, misuse detection approach is widely
adopted in the majority of commercial systems, because it is simple and 
effective, but it can not detect novel or targeted attacks. The other common
method is anomaly detection. It extracts patterns from behavioral habits of end
users, or usage history of networks and hosts. In the intrusion detection
field, supervised learning usually produces classifiers for misuse detection
from class labeled training datasets. Unsupervised learning satisfies the
requirement of anomaly detection, hence it is usually employed in anomaly
detection. The authors present two benchmarks, the DARPA-Lincoln datasets
\cite{DARPA98} and the KDD99 datasets \cite{kdd99} as the most utilized.
According to their work, the most commonly used algorithms are Neural Networks
like Feed forward Neural Networks, Radial basis function neural networks,
Recurrent Neural Networks, Self-organizing maps and Adaptive
resonance theory.

\subsection{Methods used in Grid related Intrusion Detection}
There are previously proposed methods for IDS in Grid computing. Some of them
describe schemes that are not related to Machine Learning nor computational
intelligence. For example \cite{tolba_gida:_2005} employs a relational grid
monitoring architecture, \cite{feng_ghids:_2006} presents a bottleneck
verification approach, \cite{smith_streaming_2009} describes a streaming
database approach, \cite{ungureanu_grid-aware_2010} utilizes gossip algorithms,
\cite{zhu_new_2006} represents a multi-agent approach, and 
\cite{ong_tian_choon_grid-based_2003} introduces a web services correlation
service. On the other hand, some articles are focused on ML topics such as
\cite{tolba_gida:_2005-1}, that adopts learning vector quantization Neural
Networks, \cite{schulter_intrusion_2008} and \cite{vieira_intrusion_2009}
utilize feed forward Neural Networks, \cite{jiancheng_self-adaptive_2007} 
applies auto immune systems and \cite{tolba_distributed_2005} also makes use of
learning vector quantization neural networks, all of them with a single
classifier approach. \cite{zhang_grid_2006} utilizes a hybrid approach, with a
soft computing based self-organize map dimension reduction technique, a fuzzy
Neural Network and a genetic algorithm.

\par
None of the previously presented studies applies Security by Isolation to
further improve security incident detection. Neither of them makes use of Deep
Learning approaches that allow researchers processing huge real time streams of
data produced in Grids like the WLCG. In addition, we could not find the usage
of generative methods by Recurrent Neural Networks to improve the training
datasets.

\subsection{Grid IDS related datasets}
Grid computing is a unique environment with special requirements.
We analyze the behavior of the job payloads. Therefore, using standard datasets
for ML based IDS could be ineffective. We could not find any available dataset
for IDS training in Grid computing. However several studies describe custom
metrics employed.
In \cite{tolba_gida:_2005} and \cite{smith_streaming_2009}, the authors generate
a dataset consisting of one or more log files. \cite{feng_ghids:_2006} uses an
operating system kernel module to gather system calls. In 
\cite{schulter_intrusion_2008} the measurements are extracted from audit
data of low level IDS. To identify misuse committed by insider attackers, their
system analyzes the behavior by resource usage data like CPU time and memory usage. 
The authors gather audit data from HIDS, also extracting operating system data
throughout the grid middleware and the syslog protocol. In the implementation
level they use OSSEC-HIDS and Snort. \cite{ungureanu_grid-aware_2010} uses
Snort alerts as the input metrics by the intrusion detection exchange protocol.
\cite{jiancheng_self-adaptive_2007} proposes user-level data built from user
ID, role, type and quantity of resources being consumed, and system-level data
is composed of CPU usage rate, states of main and the secondary memory and
attributes of system files. The identification, type, priority, status of
processes and the states of CPU when they are running are organized into
process-level data. IP address and port number of source and destination, type
of protocol, flags are grouped into network-level data. \cite{zhu_new_2006} 
analyses the network log data of its own monitoring area. In 
\cite{zhang_grid_2006} the extracted features are system calls (ID, return
value, return status), process (ID, IPC ID, IPC permission, exit value, exit
status) and file access (mode, path, file system, file name, argument length).
The extracted information is normalized between 0 and 1 for the input of SOM.
\cite{tolba_distributed_2005} describes a method using generated log files as a
host based intrusion detection. None of the mentioned datasets was made 
publicly available for other researchers, therefore we decided to collect our
own dataset.

\subsection{Malware Detection}
The focus of this study is to run Grid jobs securely and analyze the payloads
behavior in order to detect intrusions. Therefore, we use Linux malware samples
to test our environment and collect a dataset of malicious data. This is a more
practical approach than creating our own set of binaries with a limited set of
malicious characteristics. There are several web sites that collect malware
samples and make them available for the research community such as
\cite{virusshare} and \cite{virustotal}. In the same direction
\cite{rieck_learning_2008-1} and \cite{rieck_automatic_2011} explore Machine
Learning for malware classification, using system calls as main features for
their classifier. \cite{gibert_llaurado_convolutional_2016} explores the usage
of Deep Learning for static analysis of malware samples for classification. We
use a similar approach, however, our goal is to provide real time misbehave
detection in Grid computing, by analyzing huge flows of data that can be
generated by the millions of jobs running in the Grid.

\section{Background}
\label{background}

\subsection{Security by Isolation}
\label{isolation}
Security by Isolation is a technique that enforces component separation
(hardware or software) in a way that if one of them is compromised by an
attacker, other components still remain safe
\cite{mansfield-devine_security_2010}. There are several implementations
providing SbI such as Virtual Machines, Linux containers and the Unix multiuser
scheme. There are even security focused Operation Systems \cite{secure-os}
built with SbI as one of their core features such as \cite{Tails}, \cite{Qubes}
and \cite{SubgrahpOS}.

\subsection{Linux Containers}
\begin{figure}[h]
\begin{center}
\begin{minipage}{80mm}
\includegraphics[width=80mm]{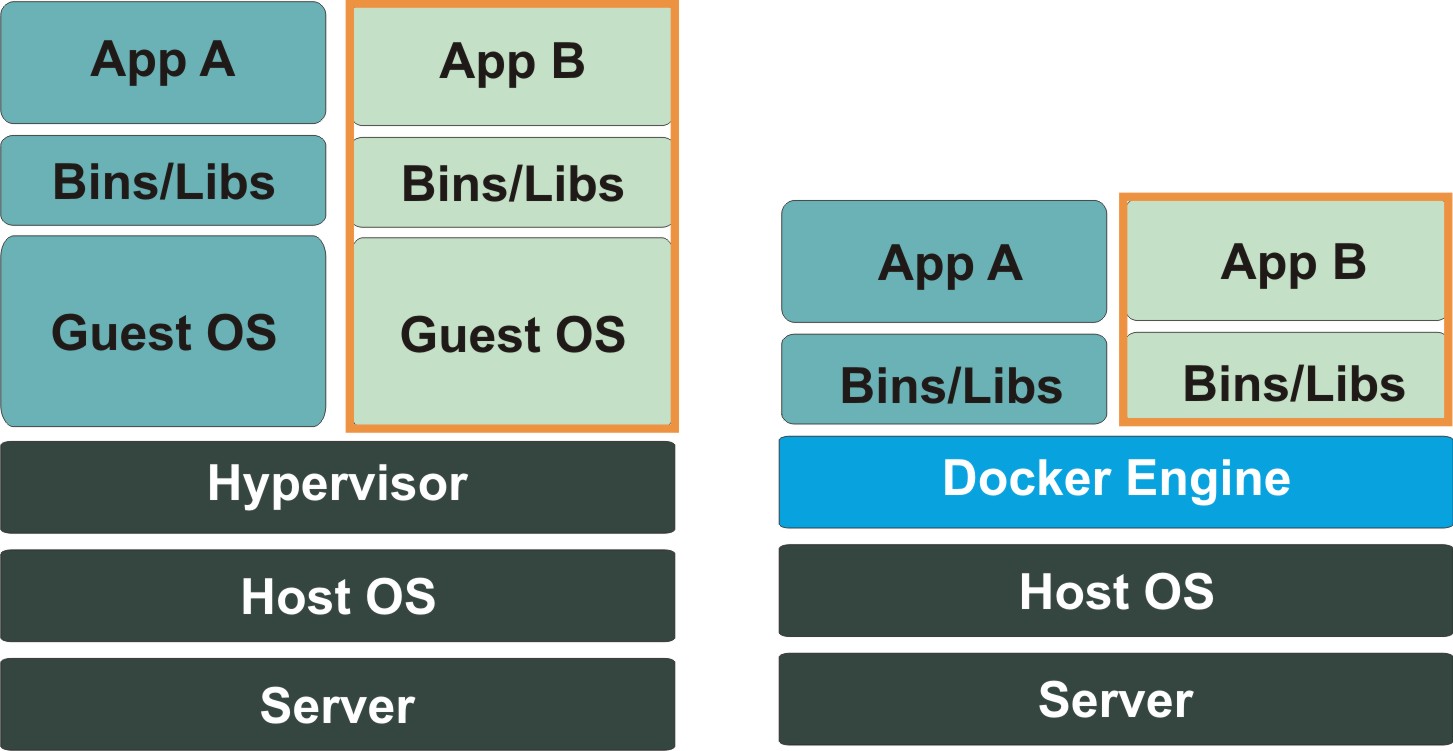}
\end{minipage}\hspace{2pc}%
\end{center}
\caption{Linux Containers on top of a common kernel}
\label{fig:containers}
\end{figure}

An LC is a set of processes running on top of a shared kernel
\cite{Xavier:2013:PEC:2497369.2497577}. They are isolated from the rest of the
machine and can not affect the host or other containers, with the exception of
exploitable vulnerabilities in the kernel or the container engine. It
takes advantage of namespaces to have a private view of the system (network
interfaces, PID tree, mount points). Cgroups are also applied to have a limited
assignment of resources. LC can be seen as an extension of the virtual memory
space concept to a wide system scope. They provide a set of features that have
advantages over other virtualization technologies. They are lightweight, fast
on booting, have a small memory footprint, and close to bare metal performance.
Figure \ref{fig:containers} describes a set containers working together,
isolated and sharing the same kernel. In opposition, Virtual Machines have
several kernels on top of a hypervisor.

\subsection{Machine Learning for data classification}
\label{classification}
ML is a form of applied mathematics \cite{Goodfellow-et-al-2016}
that tries to model how human intelligence works. One of its common application is
the statistic estimation of complicated functions. ML is frequently applied
in automatic classification of complex data, a task that traditionally has been
carried out by human operators. We can define classification in ML as follows:
the input data \(x\) is a vector of \(d\) elements \(x = (x_1 , . . . , x_d ) \in
\mathbb{R}^d\) called a feature vector. To classify the input \(x\) means to
evaluate a classification function \(C_W :\mathbb{R}_d \mapsto \{c_1 , ..., c_k
\}\) on \(x\). The output is \(c_{k^\symbol{42}} = C_W (x)\), where
\(k^\symbol{42} \in \{1 . . . k\}\); \(c_{k^\symbol{42}}\) is the class to
which \(x\) corresponds, based on the model \(W\) \cite{Bost_machinelearning}.
Different ML algorithms use different ways to find the model \(W\). We can
assume a simple two classes decision problem, defined as:
\begin{equation}
\label{eq:objective}
C(x)=W^T\phi(x)+b,
\end{equation}
where \(b\) is a bias parameter and \(\phi(x)\) is a feature-space
transformation. The training dataset corresponds to \(N\) input vectors
\(x^1,\ldots,x^N\), with target values \(c_1 ,\ldots, c_k\) where
\(c_n\in\{normal,malicious\}\). The objective is normally defined as a
loss function \(\mathcal{L}\) that represents the penalty for mismatching the
training data. The loss \(\mathcal{L}(W)\) on parameters \(W\) is the average
of the loss over the training examples \(x^1,\ldots,x^N\), as:
\begin{equation}
\label{eq:loss}
\mathcal{L}(W) = \frac{1}{N}\sum_i\mathcal{L}(W,x^i).
\end{equation}
Training consists of finding the parameters \(W\) that result in an acceptably
small loss, in the best case the smallest one (global minimum).

\subsection{Support Vector Machines (SVM)}
SVM are very popular for automated classification in Intrusion Detection
Systems \cite{tsai_intrusion_2009}. We use them to compare the classification
performance of our proposed Convolutional Neural Networks. SVM use hyperplane
decision classifiers in a similar way to traditional perceptrons. However, the
optimization objective is to maximize the margin, defined as the distance
between the decision boundary and the training data that are closest to
that hyperplane \cite{PML}. For Support Vector Machines the model \(W\) is made
of \(k\) vectors in \(\mathbb{R}^d\) where \(W=\{w_i\}_{i=1}^k\). Here our
objective is to optimize the parameters \(W\) and \(b\) such as:
\begin{equation}
\label{eq:training}
c_n(W^T\phi(x^n)+b)\geq1, n=1,\ldots,N.
\end{equation}
The optimization problem can be expressed in a simpler way as:
\begin{equation}
\label{eq:min}
\argmin_{w,b}{\frac{1}{2}\lvert\lvert W\rvert\rvert^2}.
\end{equation}

\subsection{Deep Learning}
\begin{figure*}[h]
\begin{center}
\begin{minipage}{\textwidth}
\includegraphics[width=\textwidth,scale=1.0]{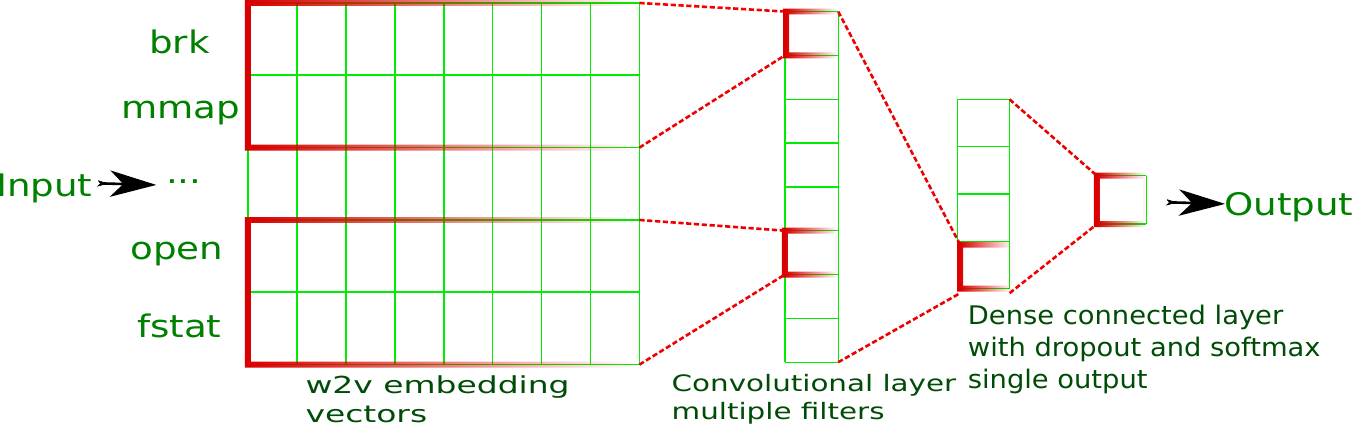}
\caption{\label{fig:cnn}A Convolutional Network for text like data processing}
\end{minipage}\hspace{1pc}%
\end{center}
\end{figure*}

DL is a sub area of Machine Learning that has solved increasingly
complicated applications with increasing accuracy \cite{Goodfellow-et-al-2016}.
Deep Learning architectures such as Convolutional Neural Networks (CNN), Deep
Belief Networks and Recurrent Neural Networks (RNN) have been utilized in
computer vision, speech recognition, natural language processing, among other
areas. They have produced some results comparable or even superior to human
experts \cite{Schmidhuber}. CNNs were first proposed by \cite{Lecun}. They are
similar to traditional Neural Networks but they use a convolution operation in
one of the layers instead of matrix operations. They are especially useful for
time series data and grid-like data topologies and have been very successful in
practical applications such as image classification and were recently proposed
for text classification \cite{convtext}. Figure \ref{fig:cnn} shows a diagram
of a Convolutional Neural Network using sliding filters to analyze text (System
calls) input data, where the convolution is applied in the first layer. A
discrete convolution operation in Deep Learning can be defined as:
\begin{equation}
\label{eq:cnn}
s(t) = (x*w)(t) = \sum_{a=-\infty}^{\infty}x(a)w(t-a),
\end{equation}
where \(x\) is an input measurement on time index \(t\), \(a\) is the age of a
measurement, \(w\) is a weighting function (also known as the kernel) that
depends on the age of measurement.

\par
Subsequent layers in CNN are normally composed of classical Deep Neural Network
full connected neurons. They are usually made of a higher amount of hidden
layers, which require special kinds of methods for updating their neuron
values. Sigmoid and rectified linear units (ReLUs) are common activation
functions \(\phi(x)\) used by CNN.

\subsection{Generative Models}
\label{generative}
Traditional Machine Learning models used for classification employ a
discriminative approach, they process input data and give a probabilistic
membership value to a certain class. On the other hand, there are ML methods
that try to learn the probability distribution function that generates the
training data (input data space)  \cite{Goodfellow-et-al-2016}. Those methods
are called generative and are useful in practical applications to create or
simulate new training data. These methods have been recently used for instance
to create new images from huge previous image datasets.

\par
Recurrent Neural Networks have been used as generative methods with important
success in applications. Long Short Term Memory (LSTM) networks were chosen in
this research. LSTM networks have an explicit memory cell and are able to
capture long-term dependences in sequential data. Formally they can be defined
as the following set of equations:
\begin{equation}
\label{eq:rnn}
{\displaystyle
{\begin{aligned}
f_{t}&=\sigma _{g}(W_{f}x_{t}+U_{f}h_{t-1}+b_{f}),\\
i_{t}&=\sigma _{g}(W_{i}x_{t}+U_{i}h_{t-1}+b_{i}),\\
o_{t}&=\sigma _{g}(W_{o}x_{t}+U_{o}h_{t-1}+b_{o}),\\
c_{t}&=f_{t}\circ c_{t-1}+i_{t}\circ
\sigma_{c}(W_{c}x_{t}+U_{c}h_{t-1}+b_{c}),\\
h_{t}&=o_{t}\circ \sigma _{h}(c_{t}),
\end{aligned}}}
\end{equation}
where \(x_{t}\) is the input vector at a given iteration \(t\), \(h_{t}\) is
an output vector, \(c_{t}\) is a cell state. \(W\) and \(U\) are parameter
matrices and \(b\) a bias vector. \(f_{t}\), \(i_{t}\) and \(o_{t}\) are gate
vectors, \(f_{t}\) is a forget gate vector, \(i_{t}\) is the input gate vector.
Finally, \(o_{t}\) is the output gate vector. Further, we give more details on
how we have adapted the LSTM network to our training data.

\section{System design}
\label{design}
In this study, we propose the integrated usage of Security by Isolation (SbI)
with Linux containers and Deep Learning methods to analyze real time monitoring
data of processes running inside virtualized HTC infrastructure, as well as the
utilization of generative methods to improve the required training data.
We introduce a hybrid supervised classification approach using word2vec for
feature selection and preprocessing and Convolutional Neural Networks (CNN) for
discrimination between normal and malicious classes. Our study also employs
Recurrent Neural Networks (RNN) for data generation on the training steps.
Arhuaco was designed as a proof of concept implementation based on this
proposed methods, with focus on Grid computing. Therefore it is designed to
provide security with a Grid based threat model approach based on the one
introduced in \cite{gomez1}. This can be applied to other types of distributed
HTC environments. In a computing Grid, an adversary may have several goals:

\begin{itemize}
  \item Steal sensitive data such as private encryption keys, user's
  certificates, tokens or credentials.
  \item Compromise user's machines to distribute malware and steal valuable 
  user information.
  \item Carry out a Denial of Service attacks.
  \item Abuse the Grid computational resources for criminal or not allowed
  activities, for instance, to deploy botnets or mining crypto-coins.
  \item Damage the organization reputation by using resources to attack other
  organizations.
\end{itemize}

\par
To achieve these goals, an attacker could use several methods:

\begin{itemize}
  \item Exploit unknown or not fixed software/hardware vulnerabilities.
  \item Listen to user network to gather sensitive clear text information.
  \item Perform a man in the middle attack.
  \item Tamper with user's jobs.
  \item Escalate privileges.
  \item Access sensitive server configuration data.
\end{itemize}

\par
In the next subsections, a detailed information about the design and
implementation of Arhuaco is given, based on our proposed contributions.
We also describe how our system works under the described threat model.

\subsection{Linux Containers for Isolation}
\begin{figure}[h]
\begin{center}
\begin{minipage}{80mm}
\includegraphics[width=80mm,scale=1.0]{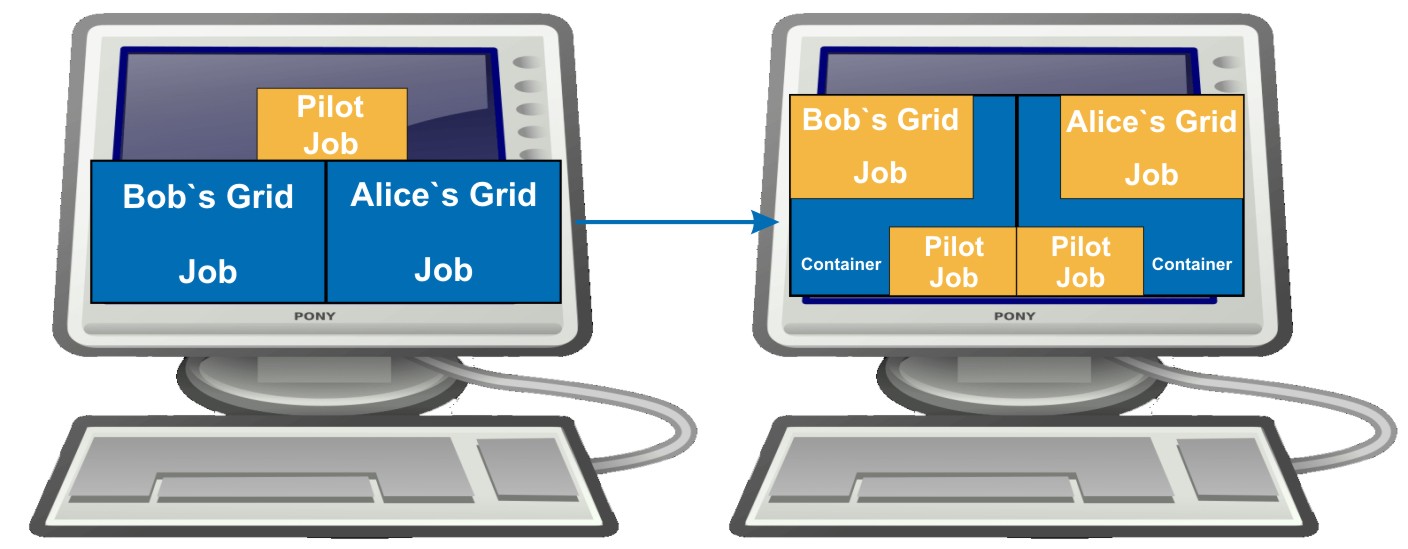}
\end{minipage}\hspace{2pc}%
\end{center}
\caption{Desired isolation scenario}
\label{fig:isolation}
\end{figure}

We require an isolation technology easily adaptable to the Linux powered Grid
computing for High Energy Physics (HEP). Linux Containers (LC) were selected to
provide SbI given their security vs performance balance \cite{ibm}. LCs provide
in addition a very important feature, network isolation. They make it possible
to create an encrypted virtual networks inside a physical or another virtual
network, in order to restrict processes running inside to access sensitive
assets. This is fundamental in Grid computing, where sites may be sharing
resources with other projects or experimental infrastructure. For instance,
there is a Grid site at the ALICE HLT cluster, sharing physical resources with
the sensitive LHC experiment network \cite{berzano_ground-up_nodate}. Therefore
virtual network isolation is used to avoid breaches. Another fundamental
feature of Linux Containers for our this study is the monitoring power they
grant. Since it is possible to encapsulate a set of processes with their own
view of the entire system, it is also possible to capture specific per process
metrics that allow us to analyze their isolated behavior. We are able to
capture resource consumption data such as CPU, memory and disk,
network connection data, and system calls for a specific container and as a
consequence for each Grid job. Therefore, we can detect with better precision
the source of a security incident or even collect forensics data for further
analysis. Figure \ref{fig:isolation} is a schema of the desired isolation
characteristics. On the left, Grid jobs run without isolation, being able to
affect other jobs or the underlying system. On the right, Grid jobs are 
isolated by LCs, each one of them run in a reduced version of the whole system,
so they have no access to other jobs or sensitive resources in the working
nodes.

\par
Traditional Grid systems use batch engines such as Condor \cite{condorgrid} for
scheduling jobs in distributed environments. To run containers instead of
standard batch jobs, we need modern orchestration tools that concede us the
ability to execute containers over a shared cluster. There are several
popular alternatives including Google Kubernetes \cite{Kubernetes}, Apache
Mesos \cite{Hindman:2011:MPF:1972457.1972488}, and Docker Swarm \cite{docker}.
In the system implementation section, we describe the reason for selecting
Docker Swarm to be the first container engine that Arhuaco interacts with.

\subsection{Deep Learning for Grid job classification}
\begin{figure}[h]
\begin{center}
\begin{minipage}{80mm}
\includegraphics[width=80mm,scale=1.0]{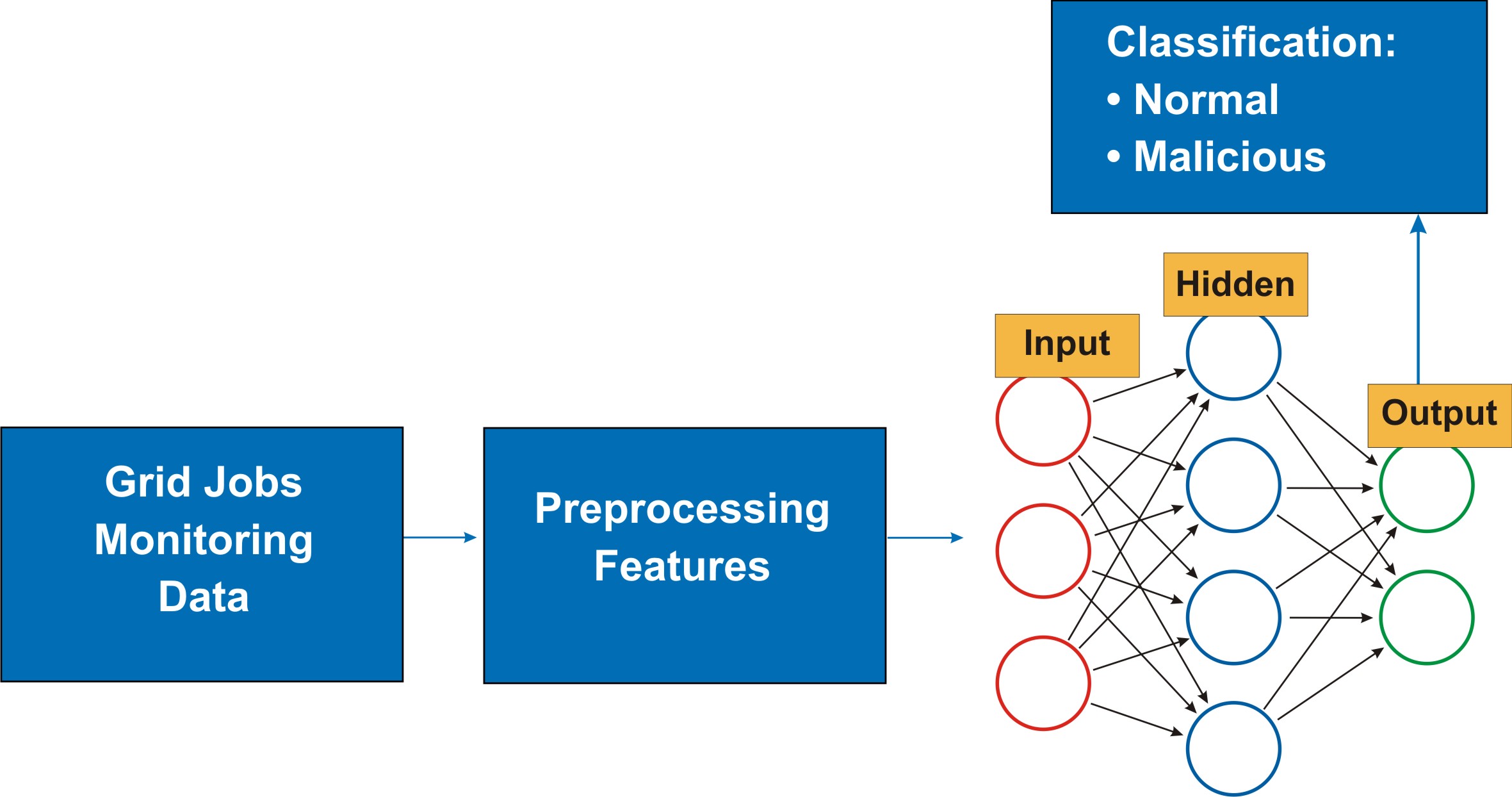}
\end{minipage}\hspace{2pc}%
\end{center}
\caption{Flow of Grid data processing for classification}
\label{flow}
\end{figure}

Popular industrial IDS such as Snort and OSSEC \cite{snort} and \cite{ossec}
use fixed rules and search for known attack signatures in order to find
possible attacks. They have problems when unknown or slightly different
intrusion methods are employed, so they need to be constantly updated
\cite{Lazarevic03acomparative}. Machine Learning has been commonly suggested in
Intrusion Detection for the modeling and analysis of log and network data for
autonomous classification of security incidents. As depicted in Figure
\ref{flow}, in this research we apply dynamic analysis of Grid jobs monitoring
data for real time intrusion detection, which means analysis of operation
system (Linux) processes. A supervised classification approach is implemented
in Arhuaco. We propose the usage of a Convolutional Neural Network architecture
based on the one introduced in \cite{convtext} for English sentences
classification and utilized in \cite{gibert_llaurado_convolutional_2016} for
static binary file classification according to their x86 machine code
instructions.

\subsection{Feature extraction}
As mentioned before, this study employs system calls and network connection
traces as input data. They are encoded in a human readable format, thus Natural
Language Processing (NLP) methods allow us to build a convenient language
model. Recently introduced Deep Learning methods for NLP implement learning
word vector representations by neural language models \cite{Bengio}. In these
vectors, words are projected from a \(1-of-V\) encoding, where \(V\) is the
number of different words available (vocabulary size), in a lower dimensional
vector space using Neural Network hidden layers \cite{convtext}. Therefore, 
semantically close words in the training corpus are mathematically close in the
lower dimensional vector space. The word2vec algorithm \cite{word2vec} was
chosen for Arhuaco to create the input features. It is a predictive model for
learning word embeddings. Word2vec vectors create suitable inputs for
Convolutional Neural Networks since they allow to treat input data as matrices,
similar to the array of pixels in an image. Here we refer to tokens instead of
words since our dataset can also contain numbers, paths, IP addresses, among
other type of data.

\par
As a preprocessing step for the traces, characters that do not increase the
amount of available information are deleted. Each trace line of system calls
is composed of the type of operation, the opcode number and all its parameters.
For network connection information each line has the DNS request, IP addresses
and ports used per connection. These lines have a variable size, therefore we
take only the first \(m\) tokens per line, adding a padding token when a line
is shorter. Then \(l\) consecutive lines are taken, which gives us a sequence
of \(n=m \times l\) total tokens, from which we extract our input feature
vector. \(a_i \in \mathbb{R}^k\) is a \(k\)-dimensional token vector
corresponding to the \(i\)-th token in our sequence. This sequence can be
described as:
\begin{equation}
\label{eq:concat}
{\displaystyle
{\begin{aligned}
a_{1:n}=a_1 \oplus a_2 \oplus \ldots \oplus a_n,
\end{aligned}}}
\end{equation}
where \(\oplus\) is the concatenation operator, and \(a_{i:i+j}\) is the 
concatenation of tokens \(a_i,a_{i+1}, . . . ,a_{i+j}\). These embedding
vectors are the result of applying word2vec method on our text input data.

\subsection{Convolutional Neural Network}
In the context of this research, a convolution operation involves a kernel or
filter \( G \in \mathbb{R}^{hk}\), which is applied to a window of \(h\) tokens
to produce a new feature. For example, a feature \(z_i\) is generated from a
window of words \(a_{i:i+h-1}\) by:
\begin{equation}
\label{eq:filter}
{\displaystyle
{\begin{aligned}
z_i=f( G \cdot a_{i:i+h-1}+b),
\end{aligned}}}
\end{equation}
where \(b \in \mathbb{R}\) is a bias term and \(f\) is a non-linear function. 
This filter is applied to each possible window of tokens in the sequence 
\(\{a_{1:h}, a_{2:h+1},..., a_{n-h+1:n}\}\) to produce a feature vector \( z= [
z_1, z_2,..., z_{n-h+1}]\), with \( z \in \mathbb{R}_{n-h+1}\). A max-over-time
pooling operation is then applied to the feature vector, taking the maximum
value \( z^*= \max{z} \) as the only feature resulting from this filter. The
most important feature, one with the highest value, is kept for each feature
map. The same process is repeated with multiple filters of different window
sizes, to obtain multiple features. These features are passed to a fully
connected \(ReLu\) layer whose output goes to the last dense layer with
\(Sigmoid\) activation with outputs corresponding to the probability
distribution over labels (normal and malicious in our problem setup).
	
\par
For the contrasted classification method - the Support Vector Machine (SVM) -
we used the well established Bag of Words (BoW) model \cite{PML} to create its
input feature vectors. In the BoW we first create a vocabulary with the list of
all possible tokens is the training set. Then we reduce this vocabulary by
using only the most used tokens. We created a vector where each component is
the number of times a given token appears in the analyzed set. An alternative
method is the Continuous Bag of Words (CBOW), which predicts target words from
source context words.

\subsection{Recurrent Neural Network (RNN) for training data generation}
A character level language model has been selected as a generative method for
Arhuaco. The objective of this model is to predict the next character in a
sequence. Given a training corpus \((c_1,...,c_T)\), where \(c_i\) is a single
character and \(T\) is the total number of characters. A Long Short Term Memory
(LSTM) RNN is utilized to determine the sequence of its output vectors
\((o_1,...,o_T)\) by a sequence of distributions \( P(c_{t+1} | c \leq t) =
\sigma(o_t)\). Here \(\sigma\) is the \(softmax\) distribution defined by:
\begin{equation}
\label{eq:sigma}
{\displaystyle
{\begin{aligned}
P(\sigma(o_t) = j) = \frac{\exp(o_{t}^{j})}{ \sum_k \exp(o^{k}_t)}.
\end{aligned}}}
\end{equation}
The objective function is to maximize the total \(\log\) probability of the
training sequence \(\sum _{t=0}^{T - 1} \log P(x_{t+1}|x \leq t)\). This
implies that the LSTM learns a probability distribution over sequences. We can
then sample from the conditional distribution  \(P(x_{t+1}|x \leq t)\) to get
the next character in a generated string and provide it as the next input to
the LSTM \cite{sutskever_generating_2011-1}. After the training process has
finished we can generate new data that can be used as extra training data in
order to extend the generalization capabilities of a classification system
\cite{gibert_llaurado_convolutional_2016}.

\subsection{Arhuaco design architecture}
\begin{figure}[h]
\begin{center}
\begin{minipage}{85mm}
\includegraphics[width=85mm,scale=1.0]{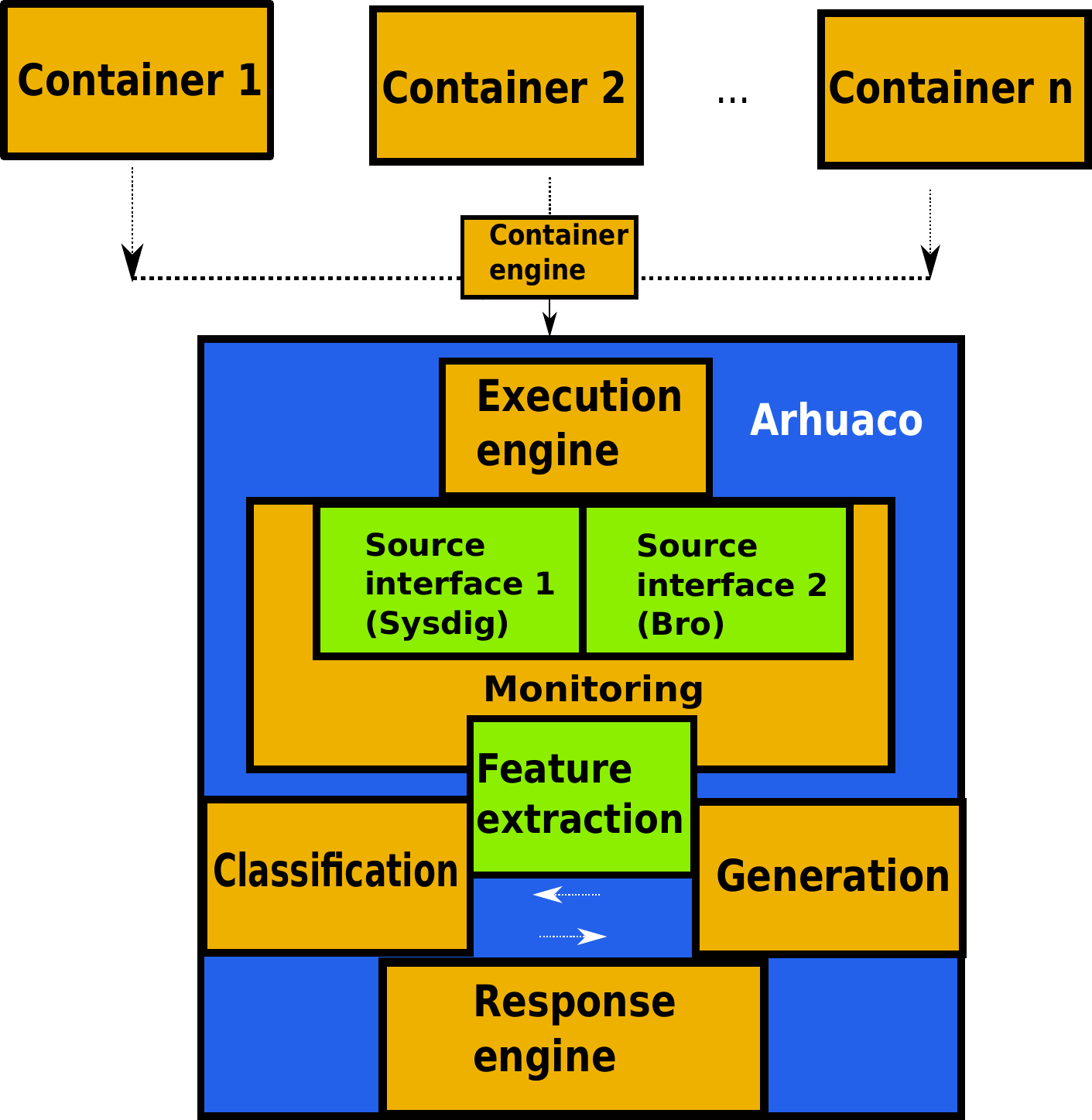}
\end{minipage}
\end{center}
\caption{Proposed Arhuaco design architecture}
\label{poc}
\end{figure}

A diagram of Arhuaco architectural components is shown in Figure \ref{poc}. The
execution engine provides an interface with a selected container scheduling
engine according to configuration parameters. Then, the executed containers are
monitored, extracting real time data for security analysis. The data is 
processed by the previously described feature extraction mechanism. Furthermore,
these preprocessed input feature vectors are sent to the classification and
generative modules. They can provide feedback to each other. Any suspicious
incident is processed by the response engine. This can be configured with
predefined actions, such as sending alerts to administrators, stopping 
suspicious jobs, or collect information for offline analysis (forensic).

\section{Implementation}
\label{implementation}
The Arhuaco prototype was developed in Python. It provides interfaces for
Grid frameworks, container engines and data collection tools. Among the
alternatives for Linux Container engines, to the most popular belong Docker
\cite{docker}, Rocket \cite{rocket}, Singularity \cite{singularity}, and LXC
\cite{lxc}. Arhuaco supports Docker in its first stage, given its broad adoption
in industrial applications and its default security measures. To be able to
execute grid jobs inside LCs, three solutions were tested: Kubernetes
\cite{Kubernetes}, Apache Mesos \cite{hindman_mesos:_2011}, and Docker Swarm.
Docker Swarm was chosen due to its simplicity and fast deployment in testing
environments. In addition, it provides out of the box encrypted virtual
networking. Further, we may create interfaces with other container scheduler
engines. A testing ALICE Grid site based on AliEn \cite{bagnasco_alien:_2008},
the ALICE Grid middelware, was deployed in a local Linux cluster at the
Goethe University in Frankfurt. It has 5 Ubuntu 14.04 nodes. A custom CentOS 6 based
Docker image and an AliEn interface for Docker Swarm were developed. CVMFS
\cite{CVMFS} was installed on the hosts and shared as a volume inside the AliEn
containers to grant access to High Energy Physcics libraries. One job per
container is executed by design, which is useful to increase the traceability
between different jobs. Besides, it is the natural micro service model for
Linux Containers.

\par
For the training and validation of our proposed classification and generative
algorithms, a dataset composed of normal and malicious system call and network
connection logs was collected. Instead of creating our own set of malicious
binary samples we have used a set of 10.000 Linux malware samples downloaded
from a security research web site \cite{virusshare}. This allows us to cover a
bigger range of malicious activities that would be very time consuming and error
prone to do manually. Regular Grid jobs were also collected from the ALICE Grid
production environment, using our test Frankfurt site, to improve the training
data. We ran the samples and collected the same set of metrics for both types
of binaries. We executed them inside containers and used the isolation and
monitoring features to collect every system call and network connection. These
are some examples of the collected data:

\begin{verbatim}
Malware:  
* IP.x IP.y irc.qeast.net 1 C_INTERNET ...
* file open fd 4 name /etc/passwd ...
Grid job:
* IP.z IP.w alice-disk-se.gridka.de 1 ...
* file access res 2 ENOENT 
  name /cvmfs/alice.cern.ch/x86 ...
\end{verbatim}

\par
We utilized Sysdig \cite{sysdig} for collecting system calls and Bro IDS
\cite{BRO} for network connection data. For executing the malware samples a
testing environment without Internet access was deployed, using Inetsim
\cite{Inetsim} for network connection emulation. We have also employed Cuckoo
sandbox \cite{Cuckoo} to isolate and monitor these runs. Table \ref{tablefull}
shows a summary of the collected dataset. In the Arhuaco online setup, once the
training is finished, the system call and network traces collection is done in
real time, as well as the classification.

\begin{table}
\caption{\label{tablefull}Full preprocessed available datasets}
\begin{center}
  \begin{tabular}{ l p{0.2\linewidth}cc }
    \hline\noalign{\smallskip}
    Dataset & Normal & Malware \\
    \noalign{\smallskip}\hline\noalign{\smallskip}
    System call & 12GB 127'100.000 lines & 8.2GB - 127'054.763 lines
    \\
    Network & 868KB 20.733 lines & 108KB - 2.937 \\
    \noalign{\smallskip}\hline
  \end{tabular}
\end{center}
\end{table}

\begin{table}
\caption{\label{tableshort}Used training and validation data after feature 
extraction}
\begin{center}
  \begin{tabular}{ lcc }
    \hline\noalign{\smallskip}
    Dataset & Training & Validation \\
    \noalign{\smallskip}\hline\noalign{\smallskip}
    System calls traces & 10'000.000 & 100.000 \\
    Network traces & 20.000 & 2.000 \\
    \noalign{\smallskip}\hline
  \end{tabular}
\end{center}
\end{table}

\begin{figure}[h]
\begin{center}
\begin{minipage}{80mm}
\includegraphics[width=80mm,scale=1.0]{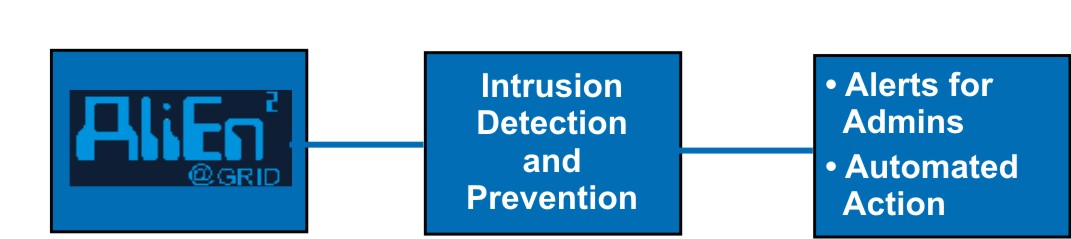}
\end{minipage}\hspace{2pc}%
\end{center}
\caption{Proof of concept diagram in the ALICE GRID}
\label{figureids}
\end{figure}

Table \ref{tableshort} shows a summary of the obtained samples after the
feature selection step. A representation of the implemented test as part of the
ALICE Grid can be seen in Figure \ref{figureids}.

\par
We have implemented the Deep Convolutional Neural Network (CNN), Support Vector
Machine (SVM) and Recurrent LSTM Neural Network by using the Python library 
Keras \cite{chollet2015keras} with Theano \cite{theano} as a backend. Keras
simplifies the development of Deep Learning algorithms and provides parallel
computing capabilities powered by Theano. It also supports TensorFlow
\cite{TensorFlow}. It is convenient for High-throughput Computing
(HTC) applications, since it can share resources with other applications
running in parallel, and it is strongly focused on using GPU to increase the
parallel processing performance. Another Python library, Gensim
\cite{rehurek_lrec} was employed for word2vec extraction of embedding vectors.
Due to the huge size of the training corpus, it has the functionality to deal
with data that does not fit in main memory.

\par
The CNN hyper-parameters have been selected by an empiric grid search. They are
listed in Table \ref{cnnpar}. We use momentum and parameter decay to ensure
the model convergence, and dropout to prevent overfitting. For the Support
Vector Machine, we chose the \(Hinge\) loss function and the \(Adadelta\)
optimizer. The parameters \(m\), \(l\) and \(n\) described in the previous
section are shared among the two models. They are listed in Table 
\ref{sharedpar}. These last parameters were selected to keep a good balance
between the classification accuracy of normal and malicious classes and the
ability to detect intrusions in real time. For the LSTM Network, the chosen
optimizer is Root Mean Square Propagation, with a learning rate of 0.01, and a
categorical cross entropy as loss function.

\begin{table}
\caption{\label{cnnpar}Custom CNN parameters}
\begin{center}
  \begin{tabular}{ lcc }
    \hline\noalign{\smallskip}
    Parameter & System calls & Network\\
    \noalign{\smallskip}\hline\noalign{\smallskip}
    Embedding dimension & 20 & 10 \\
    Filter sizes & 3, 4, 5 &  2, 3\\
    Total number filters & 20 & 3 \\
    \noalign{\smallskip}\hline\noalign{\smallskip}
    Optimization function & \multicolumn{2}{c}{SGD} \\
    Learning rate & \multicolumn{2}{c}{0.001} \\
    Momentum & \multicolumn{2}{c}{0.80} \\
    Decay & \multicolumn{2}{c}{\(10e-6\)} \\
    \noalign{\smallskip}\hline
  \end{tabular}
\end{center}
\end{table}

\begin{table}
\caption{\label{sharedpar}Shared SVM and CNN parameters}
\begin{center}
  \begin{tabular}{ lccc }
    \hline\noalign{\smallskip}
    Dataset & \(m\) & \(l\) & \(n\) \\
    \noalign{\smallskip}\hline\noalign{\smallskip}
    System call & 7 & 6 & 42 \\
    Network & 5 &  1 & 5\\
    \noalign{\smallskip}\hline
  \end{tabular}
\end{center}
\end{table}

\section{Evaluation}
\label{results}
In this section, an empiric evaluation of the performance impact, classification
ability and data generation effectiveness for the methods implemented in
Arhuaco is described. Here the goal was to answer the following questions:

\begin{itemize}
  \item Does using LC for isolation and system call monitoring create a
big performance impact? If that is the case, is it critical in
comparison to the increment in security provided?
  \item Does the CNN for classification of job traces provide better accuracy
  and false positives rates than the traditional SVM?
  \item Does the LSTM Network for data generation improve the training
  results?
\end{itemize}

\subsection{Evaluation methods}
We have deployed two type of performance tests. The first with the Linpack
benchmark as introduced in \cite{ibm} to test the throughput. The second
measures the performance impact in processing times. For the first test,
we ran Linpack jobs while for the secong the job is a production ALICE script.
We have prepared 3 evaluation scenarios: running jobs on a Linux machine, then
in Docker containers and finally in Docker containers with system call
interception and processing by Arhuaco. In our performance analysis, the impact
of collecting network data from the job is not measured.

\par
Loss, Accuracy (ACC) and False Positive Rate (FPR) are the measurements for the
correctness of the compared classification methods. The overall performance of the trained
classifiers is evaluated by the Accuracy. This measures the number of instances
that were correct, which are both the True Positives and True Negatives, over
the entire size of the dataset, which is the number of True Positives (TP),
True Negatives (TN), False Positives (FP), and False Negatives (FN). It is
calculated as:
\begin{equation}
\label{eq:acc}
Accuracy (ACC) = \frac{(TP+TN)}{(TP+TN+FP+FN)},
\end{equation}
\begin{equation}
\label{eq:fpr}
False Positive Rate (FPR) = \frac{FP}{FP+TN},
\end{equation}
where False Positives (FP) is the number of misclassified instances of a
certain category over the number of all instances that are classified as that 
category. The ACC and FPR values are defined in the range \([0,1]\), where \(1\)
is the best possible value for ACC and the worst for FPR. On the other hand,
\(0\) is the best value for FPR and the worst for ACC. They can also be
interpreted as percentage values in the range \([0\%,100\%]\).

\subsection{Performance impact results}
The performance evaluations were carried out in our Frankfurt testing Grid
site. Each of the 5 machines has identical configuration: 4 Intel Xeon (64 bit)
processors for a total 16 cores running at 2.27 GHz and 16 GB RAM. They all
have Ubuntu 14.04 as the installed operating system. The container images are
derived from CentOS 6, since this is the current recommended distribution to be
used with the CERN HEP Grid libraries.

\begin{table}
\caption{\label{Linpack}Performance impact in throughput measurement}
\begin{center}
  \begin{tabular}{ lc }
    \hline\noalign{\smallskip}s
    Setup & Linpack (GFLOPS)\\
    \noalign{\smallskip}\hline\noalign{\smallskip}
    Native & 3.7664 \\
    Docker & 3.7506 (-0.4\%) \\
    Docker plus sysdig &  3.7463 (-0.5\%) \\
    \noalign{\smallskip}\hline
  \end{tabular}
\end{center}
\end{table}

\par
\begin{figure}[h]
\begin{center}
\begin{minipage}{90mm}
\includegraphics[width=90mm,scale=1.0]{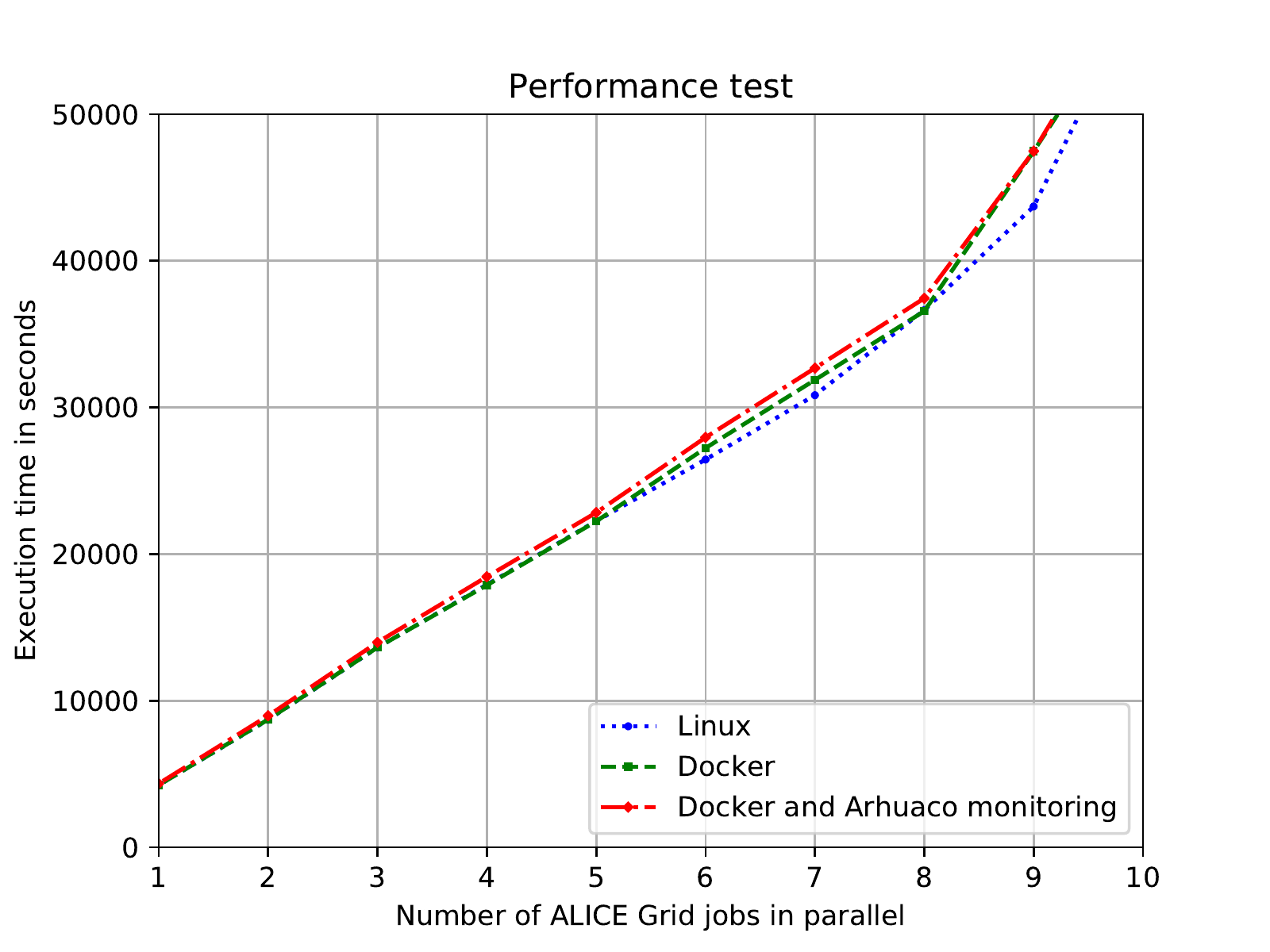}
\end{minipage}\hspace{2pc}%
\end{center}
\caption{Performance impact comparison}
\label{perf}
\end{figure}

Table \ref{Linpack} shows the performance of Linpack jobs on Linux, Docker and
Docker plus Arhuaco monitoring. We have executed one job per CPU core, for a
total of 16 jobs and have measured the resulting throughput. Figure \ref{perf}
describes the results of the ALICE Grid job test. Up to 10 jobs instances in
parallel were executed. The measurement was the execution time they spend to
be completed. Each job does the same tasks, simulation, reconstruction and
analysis of the same ALICE HEP data.

\par
The resulting throughput is almost identical for the 3 measured scenarios. This
can be explained by the little involvement the containers and system call
monitoring have in the floating point operations. For the execution time
tests with ALICE jobs, there was a maximum performance impact of 8.634\% when
using Docker in comparison to the Linux execution. This can be seen in the
graphic when the number of ALICE jobs in parallel is equal to 9. For the Docker
plus Arhuaco monitoring runs, the maximum performance impact was of 2.535\% in
comparison to the Docker runs (when the number of parallel jobs are equal to 7).
It is noticeable that there is not a big difference in using only Docker containers
in comparison to Docker and Arhuaco monitoring. To further reduce the impact on
the Grid jobs performance of the system call monitoring method, we provide
several configuration options: the first is to analyze only a set of random
jobs. The second option is to first analyze only network data and then in case of
suspicious behavior activate the system calls collection. The third possibility
is a secure mode were all the data is analyzed, which is the default mode.

\subsection{Supervised classification results}
\begin{figure*}[!htb]
\begin{center}
\begin{minipage}{0.49\textwidth}
\includegraphics[width=\linewidth]{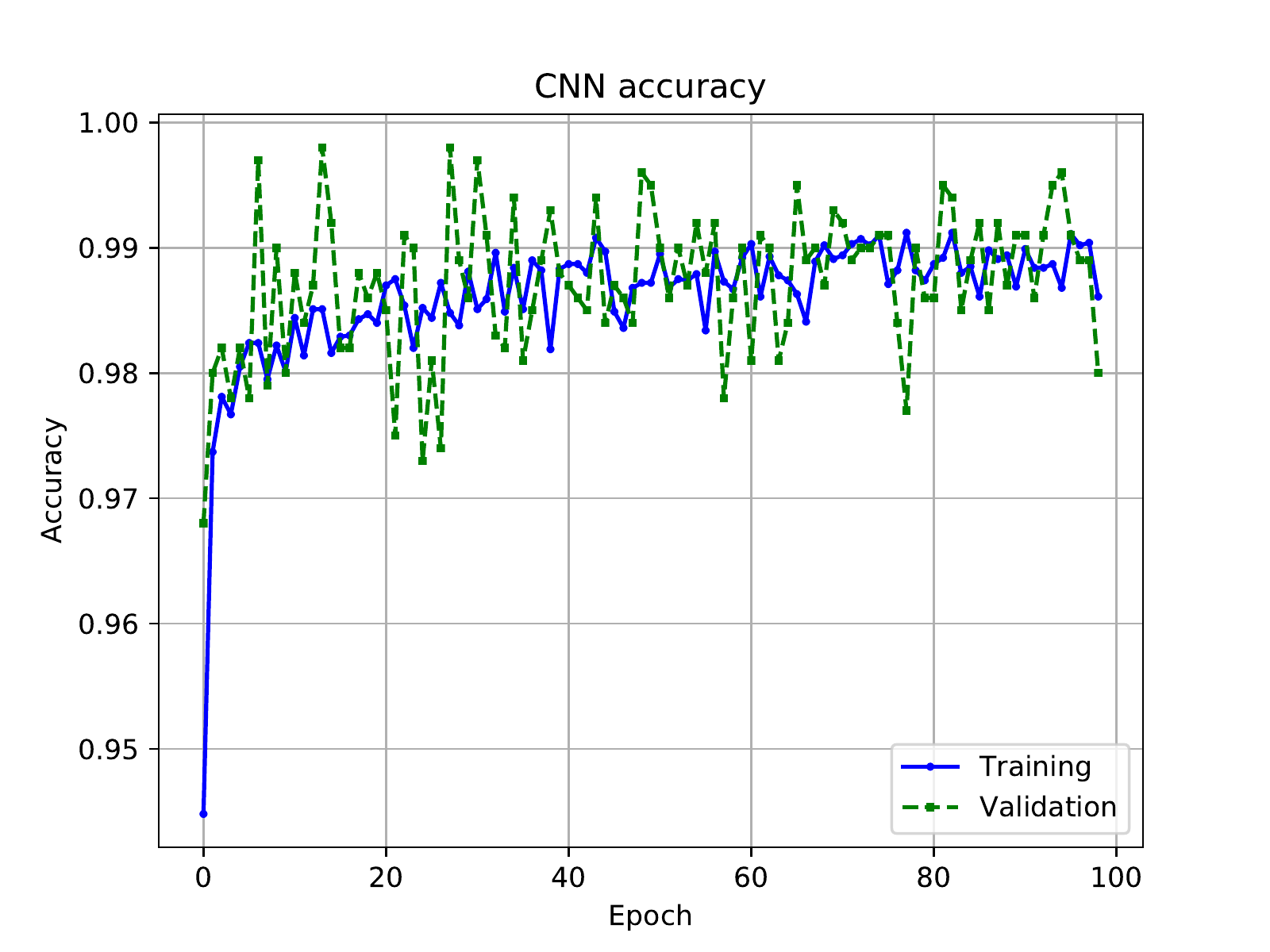}
\caption{\label{fig:sys_conv_accuracy}Classification accuracy of the CNN
applied to system calls}
\end{minipage}
\begin{minipage}{0.49\textwidth}
\includegraphics[width=\linewidth]{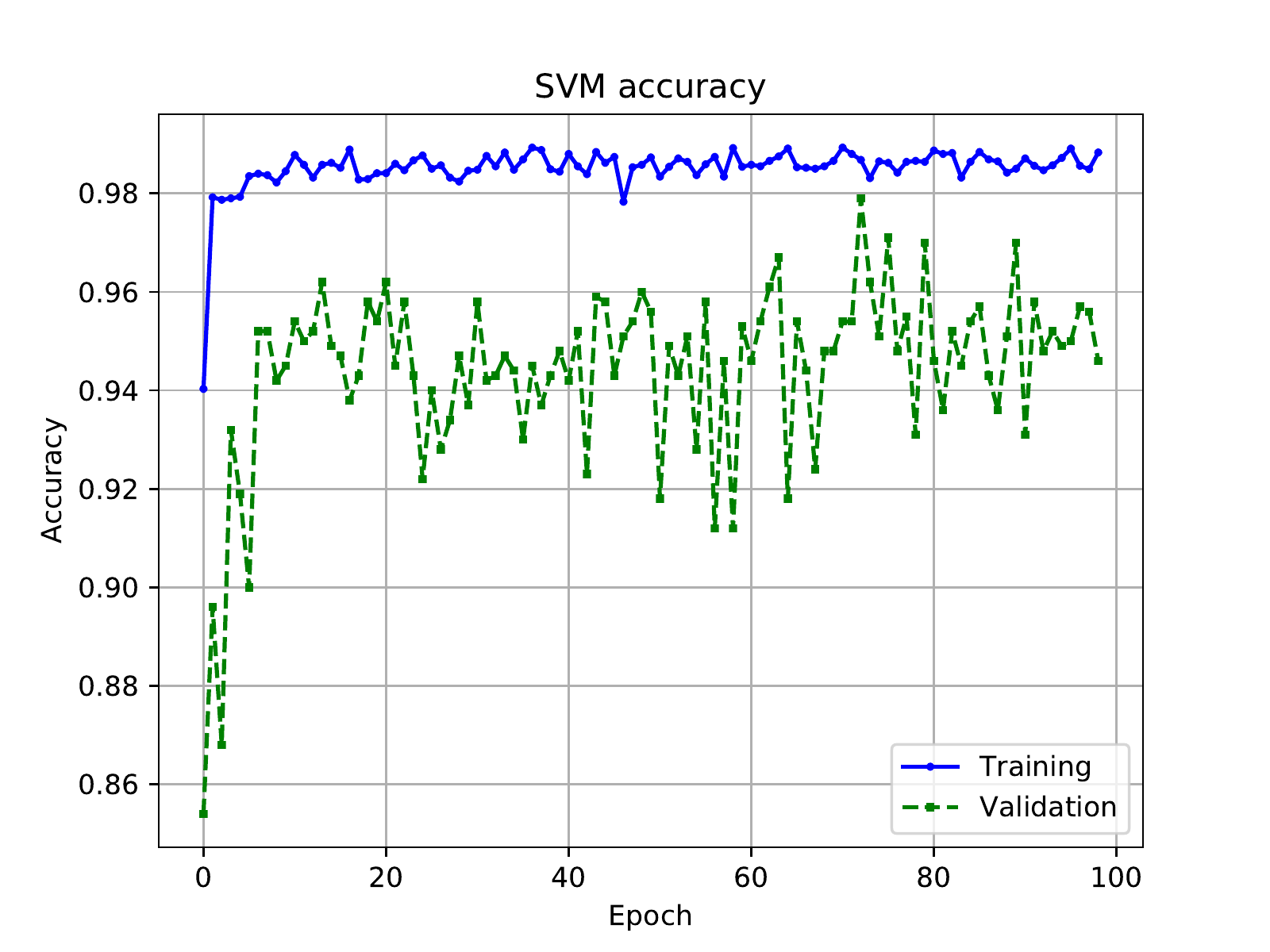}
\caption{\label{fig:sys_svm_accuracy} Classification accuracy of SVM
applied to system calls}
\end{minipage}
\end{center}
\end{figure*}

\begin{figure*}[!htb]
\begin{center}
\begin{minipage}{0.49\textwidth}
\includegraphics[width=\linewidth]{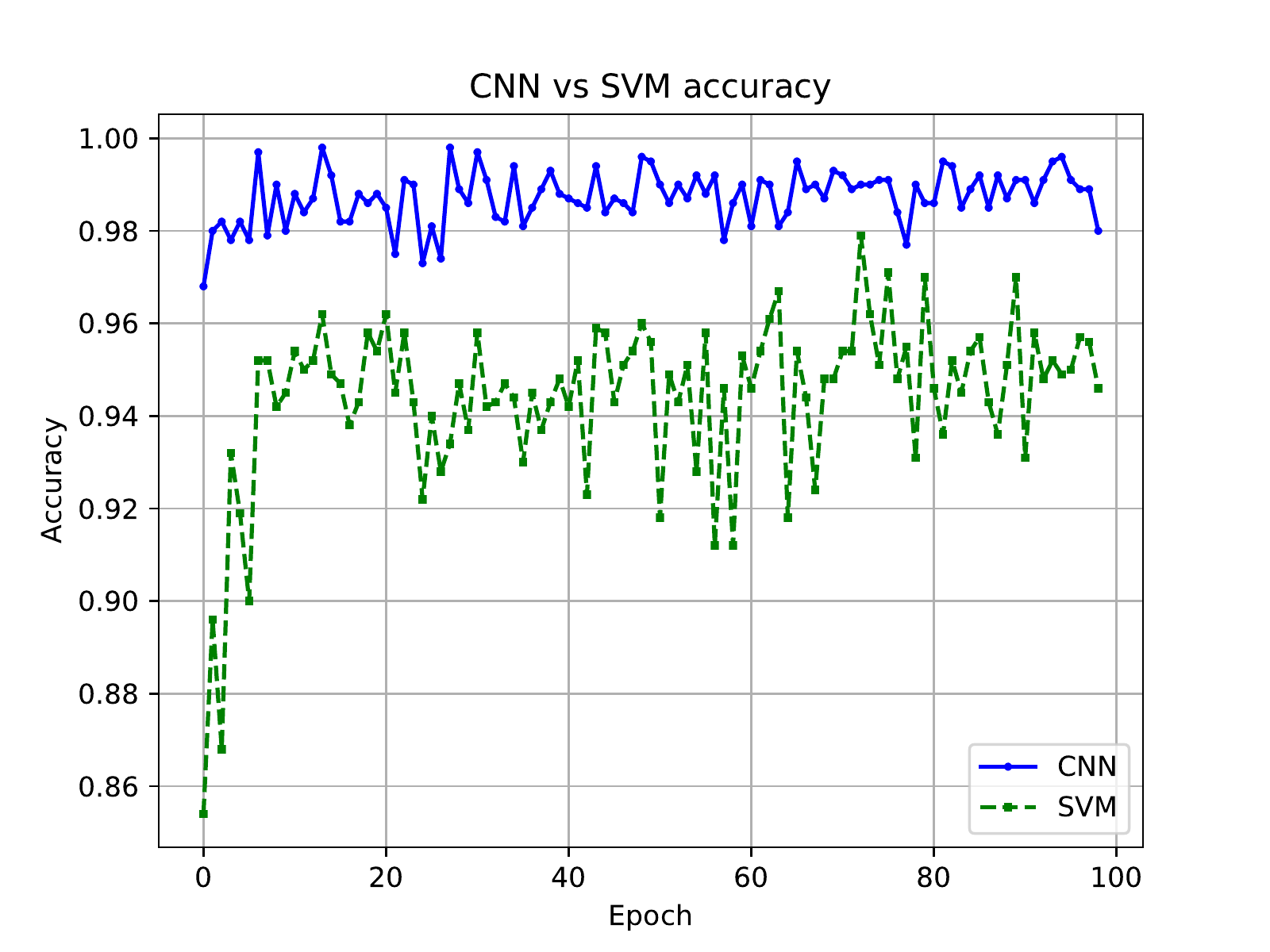}
\caption{\label{fig:sys_cnn_svm_acc} Accuracy comparison of CNN vs SVM
applied to system calls }
\end{minipage}
\begin{minipage}{0.49\textwidth}
\includegraphics[width=\linewidth]{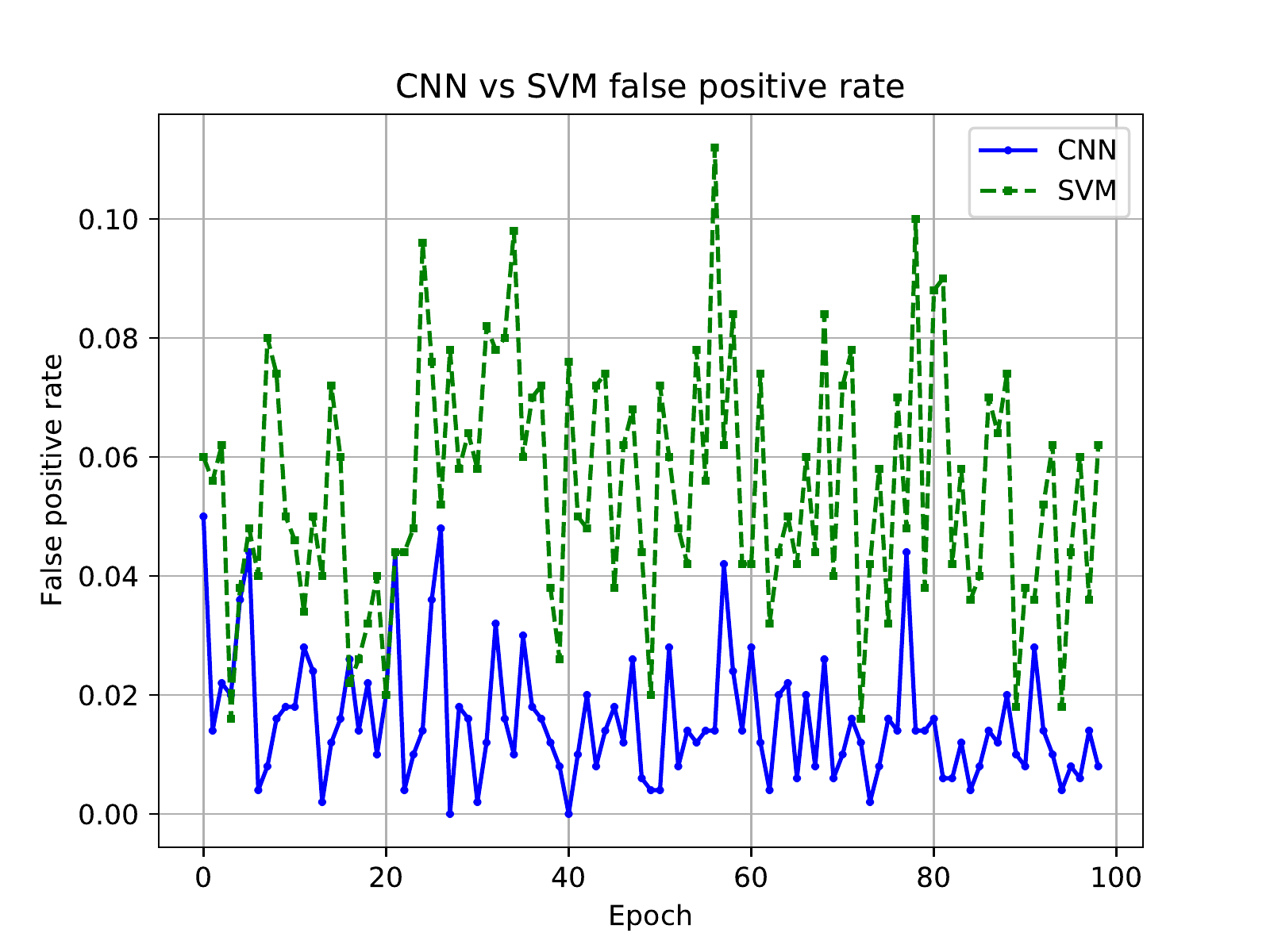}
\caption{\label{fig:sys_cnn_svm_fpr} False
Positive Rate comparison of CNN vs SVM applied to system calls }
\end{minipage}
\end{center}
\end{figure*}

Figure \ref{fig:sys_conv_accuracy} and Figure \ref{fig:sys_svm_accuracy}
indicate the training and validation accuracy for the proposed Convolutional
Neural Network (CNN) and the compared Support Vector Machine (SVM) on the
analysis of system call traces, preprocessed by word2vec and BoF. The CNN
validation curve is close to the training curve, approaching nearly 99\%. Since
the SVM model is far smaller, a trend for over-fitting can be seen in the
validation curve compared to the training curve. Figure
\ref{fig:sys_cnn_svm_acc} and Figure \ref{fig:sys_cnn_svm_fpr} show a
comparison for the ACC and FPR with the validation data. Clearly, the CNN
provides a higher accuracy and lower False Positive Rate than the SVM.

\par
\begin{figure*}[!htb]
\begin{center}
\begin{minipage}{0.49\textwidth}
\includegraphics[width=\linewidth]{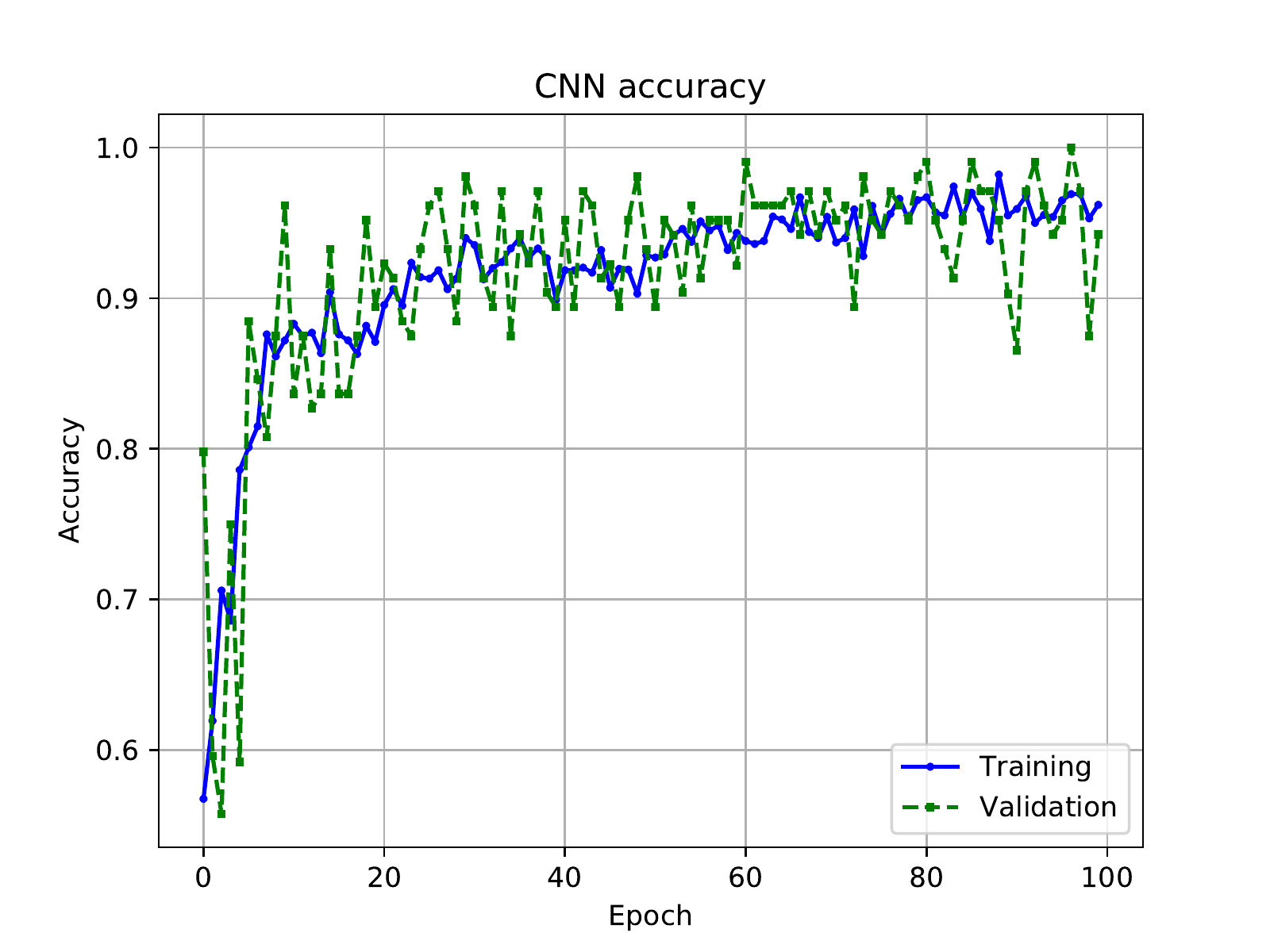}
\caption{\label{fig:net_conv_acc} Classification accuracy of CNN applied
network data}
\end{minipage}
\begin{minipage}{0.49\textwidth}
\includegraphics[width=\linewidth]{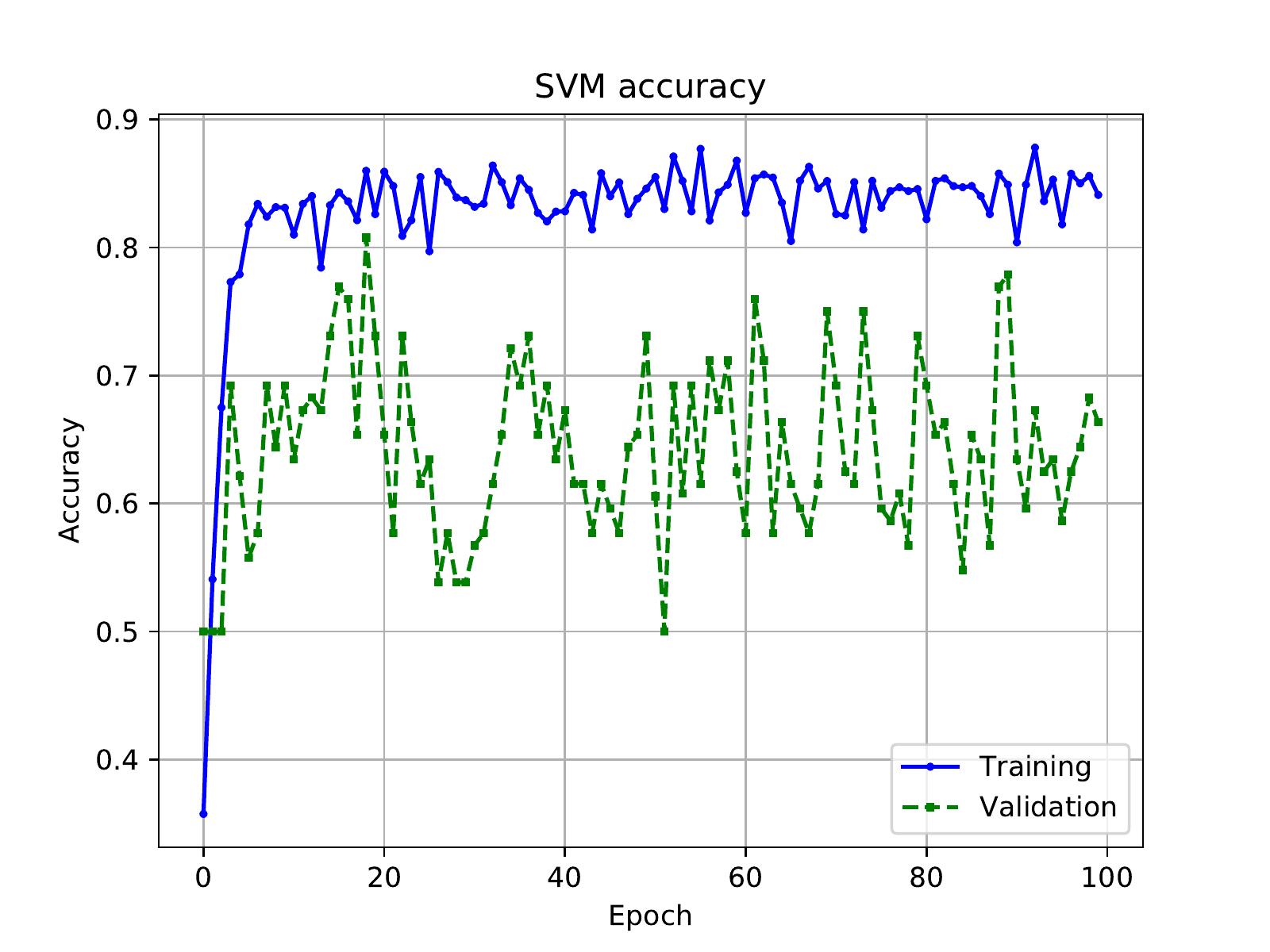}
\caption{\label{fig:net_svm_acc} Classification accuracy of SVM applied
to network data}
\end{minipage}
\end{center}
\end{figure*}
\begin{figure*}[!htb]
\begin{center}
\begin{minipage}{0.49\textwidth}
\includegraphics[width=\linewidth]{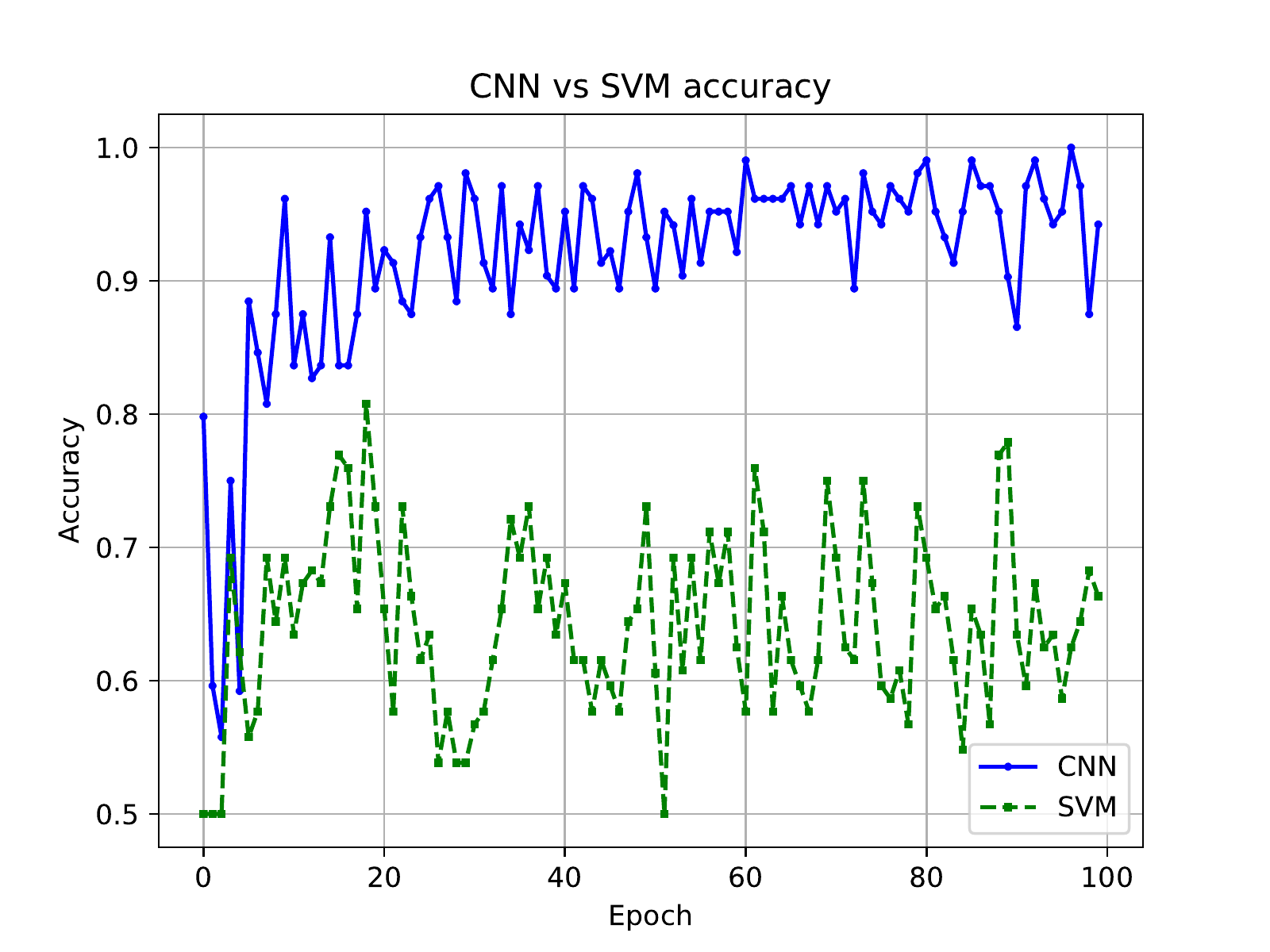}
\caption{\label{fig:net_cnn_svm_acc} Classification accuracy
comparison of CNN vs SVM applied to network data}
\end{minipage}
\begin{minipage}{0.49\textwidth}
\includegraphics[width=\linewidth]{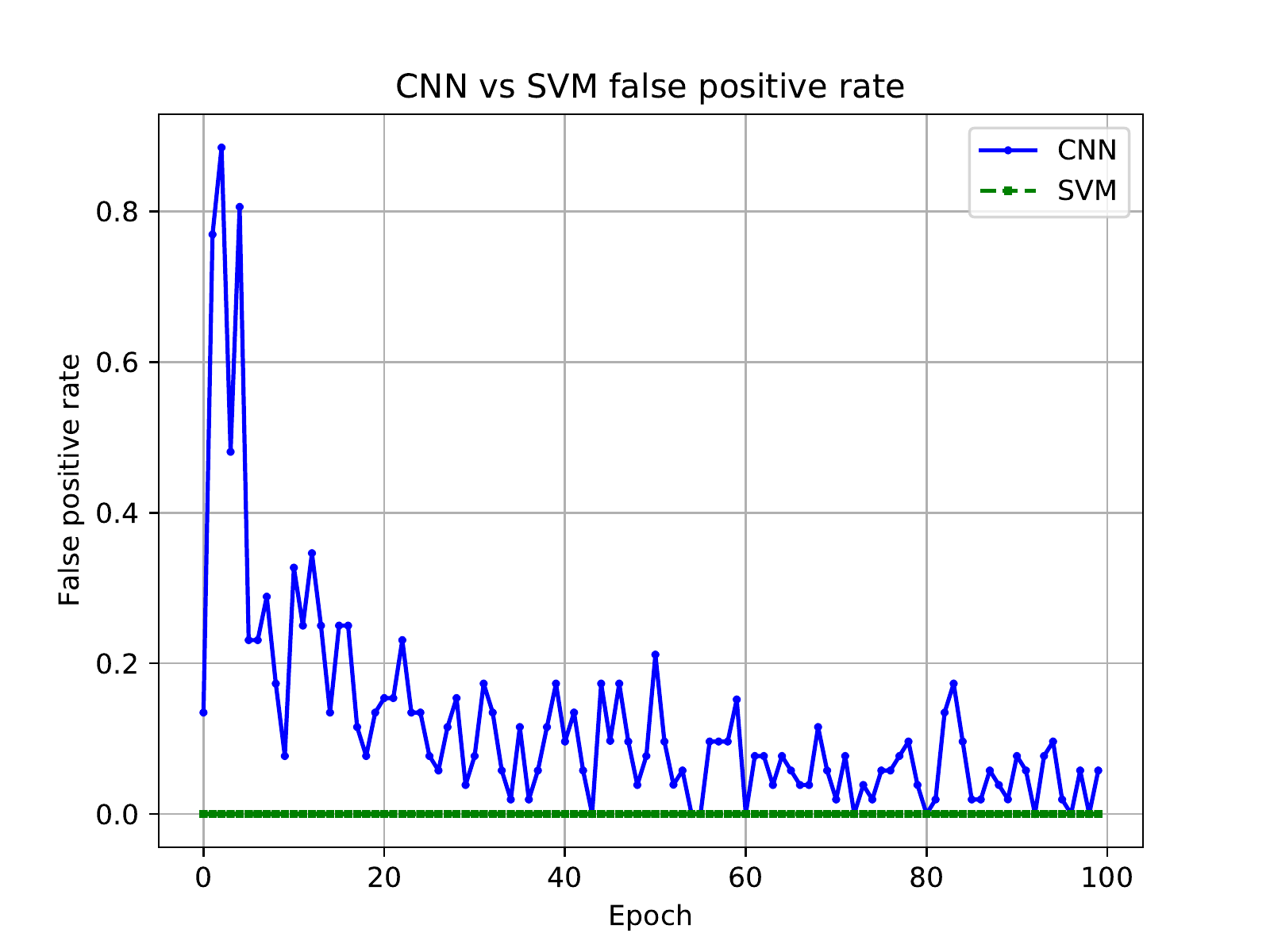}
\caption{\label{fig:net_cnn_svm_fpr} Classification False
Positive Rate of CNN vs SVM applied to network data}
\end{minipage}
\end{center}
\end{figure*}

Figure \ref{fig:net_conv_acc} and Figure \ref{fig:net_svm_acc} represent the
the results for the classification of network traces with our proposed CNN and
compared SVM method. As with the system calls case, the CNN shows similar
results with the training and validation data, approaching 99\%. The SVM
shows signs of over-fitting and a lower accuracy. In Figures
\ref{fig:net_cnn_svm_acc} and \ref{fig:net_cnn_svm_fpr} we see the comparison
of the validation data for accuracy and False Positive Rate. A better accuracy
for the CNN is clear in this case as well, however, the SVM exhibits a better
FPR.

\par
Table \ref{results} summarizes the results obtained for applying the
Convolutional Neural Network and Support Vector Machines on previously
unseen testing data. A better accuracy and False Positive Rates for the CNN
over the SVM is shown, except in the case of FPR for the network data. This
validates the results we have obtained in the training steps and demonstrates
the effectivenesses of using Convolutional Neural Networks for classification in
Arhauco to detect intrusions, in comparison to utilizing Support Vector
Machines, the most popular method used in Grid computing Intrusion Detection
Systems.

\begin{table}
\caption{\label{results} Classification accuracy and FPR comparison of testing
data}
\begin{center}
  \begin{tabular}{ lcccc }
    \hline\noalign{\smallskip}
    \makecell{Testing \\ Dataset} & CNN Acc. & SVM Acc. & CNN FPR & SVM FPR\\
    \noalign{\smallskip}\hline\noalign{\smallskip}
    \makecell{System \\ call} & 0.9657 & 0.9499 & 0.0632 & 0.0839 \\
    \makecell{Network \\ traces} & 0.9733 & 0.8011 & 0.04004 & 0.0 \\
    \noalign{\smallskip}\hline
  \end{tabular}
\end{center}
\end{table}

\subsection{Generative model results}
\begin{figure*}[!htb]
\begin{center}
\begin{minipage}{0.49\textwidth}
\includegraphics[width=\linewidth]{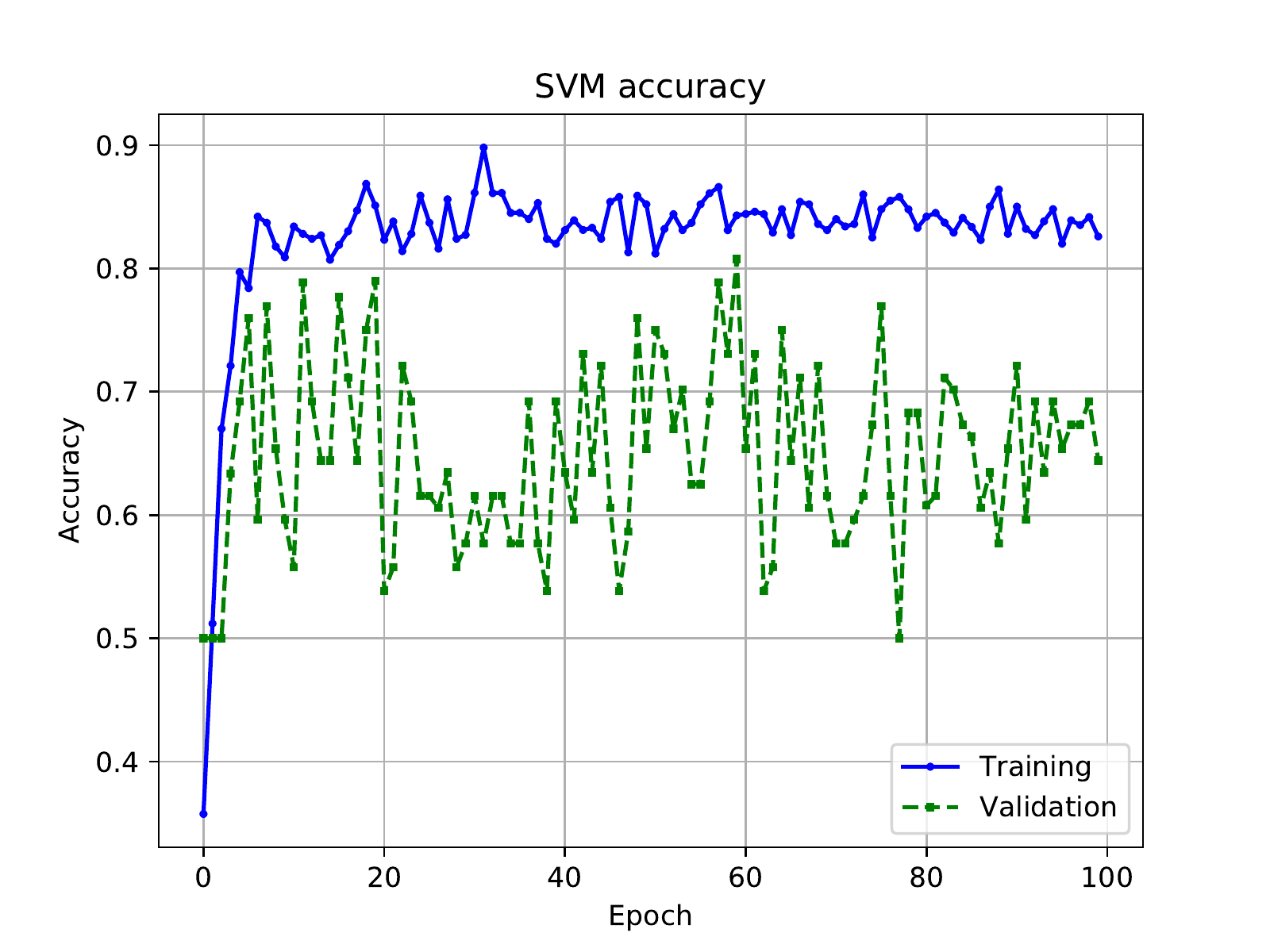}
\caption{\label{fig:net_svm_accuracy-generated} Classification
accuracy of SVM applied to network data with the newly generated data for
training}
\end{minipage}
\begin{minipage}{0.49\textwidth}
\includegraphics[width=\linewidth]{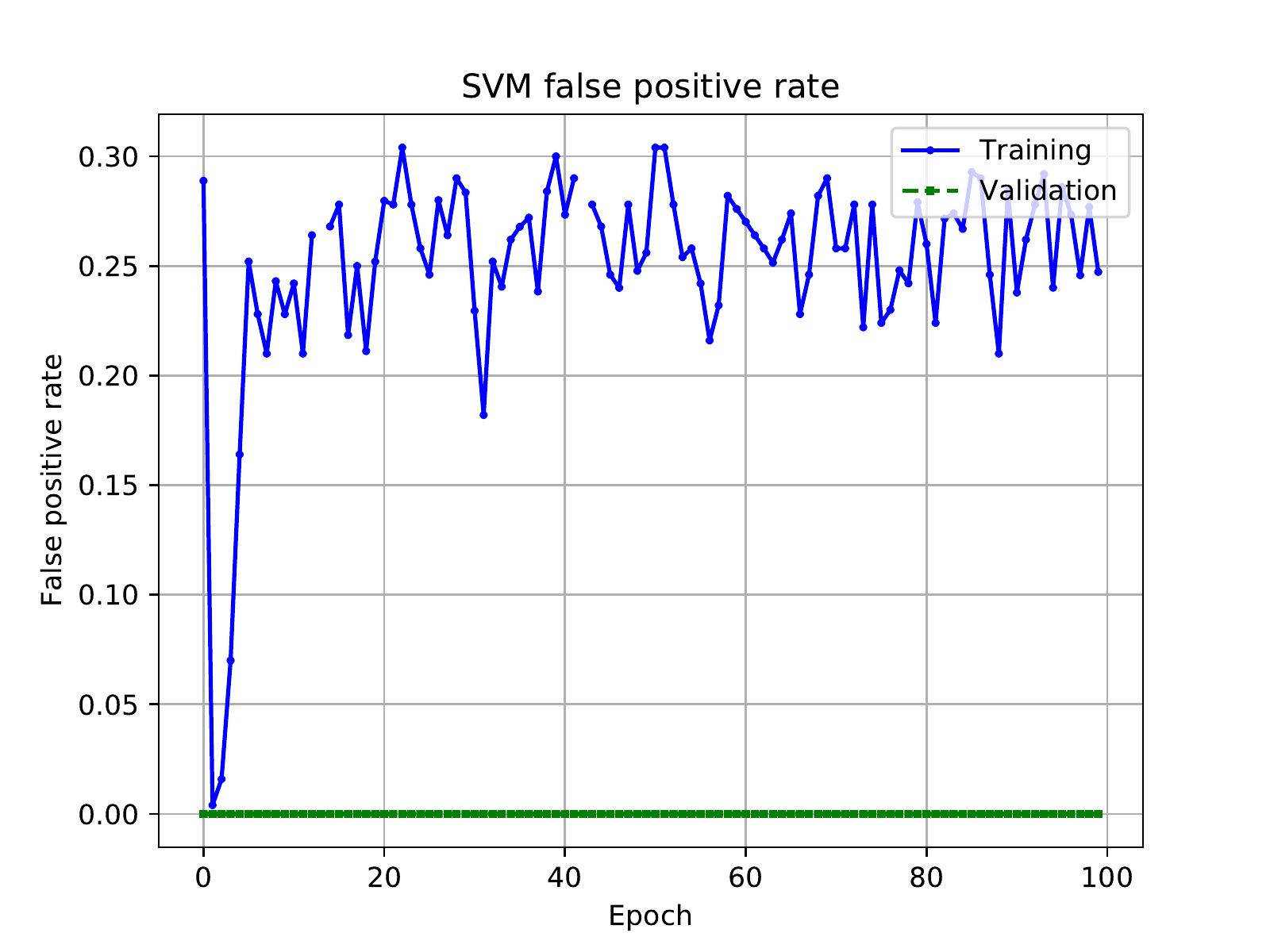}
\caption{\label{fig:net_svm_fpr-generated} Classification
False Positive Rate of SVM applied to network data with the newly generated data
for training}
\end{minipage}
\end{center}
\end{figure*}

As shown in Table \ref{results}, the SVM classification results for network
connection information produced the worst accuracy in all our tests. Therefore
we utilized this same case to prove the effectiveness of the proposed generative
technique. The Deep RNN was trained with the available network data corpus and
then modified to generate 20\% new training data. The new data was concatenated
to the original data to create a new training set. Further, the SVM is trained
again to measure the new accuracy. The validation data utilized belongs to the
original data. Figures \ref{fig:net_svm_accuracy-generated} and 
\ref{fig:net_svm_fpr-generated} provide a new comparison of the SVM accuracy
and FPR after being trained with the additionally generated data. A noticeable
improvement can be seen regarding the training with the original data in
Figures \ref{fig:net_cnn_svm_acc} and \ref{fig:net_cnn_svm_fpr}. This
demonstrates the practical benefits of using LSTM for modeling and generating
data in the context of Intrusion Detection Systems.

\par
The obtained accuracy for the SVM trained with the newly generated data, applied
to unseen network data from the original dataset was 0.8201. There was a 2.38\%
improvement rate as expected with our approach, in comparison the original value
from Table \ref{results}, 0.8011. This validates the results obtained in the
training steps.

\section{Discusion and future work}
\label{discusion}
Regarding the first question we have defined in our evaluation, the performance
tests present an execution time overhead of 0.4\% when using Linux Containers in
comparison with the Linux runs for the Linpack throughput test, and 0.5\% when
using Arhuaco monitoring. This can be considered as a very small impact. The
impact is more considerable when testing the execution time for the ALICE Grid
job. The Docker container generates up to 8.634\% of overhead. It is higher, an
extra 2.535\% when monitoring by Arhuaco is added on top. Although this
overhead is not critical, we have implemented some ideas to reduce it. We have
a two layers approach, based on user configuration. Arhuaco can detect
malicious activities by first intercepting and analyzing network connections
and then it can make a deeper analysis by a second layer using the system
calls, after suspicious processes detection. Another configurable option is to
randomly analyze a small set of jobs running in the Grid, which still can
contribute to have an improved level of security. It is important to notice
that containers are being increasingly utilized in Grid computing
collaborations and as we have shown, adding extra monitoring and analysis by
Arhuaco creates an acceptable performance impact. Another point to
consider is that there are studies \cite{ibm} that compare Linux Containers
against Virtual Machines (VMs). They demonstrate that VMs create a bigger
overhead in the performance than LCs.

\par
The classification algorithm implemented in Arhuaco improves the detection of
malicious activities running inside the grid, compared to traditionally
employed methods, which responds to the second research question. Convolutional
Neural Networks demonstrate a better classification ability than Support Vector
Machines for system calls as well as for network connections in Grid jobs. Since
our gathered malicious network data is rather small, we have successfully
generated new data by a Recurrent Neural Network. We described how it allowed us to
improve the classification results for the SVM, giving an answer to the third
evaluation question. An interesting point to remark is that our approach of
analyzing input as a text data by Natural Language Processing approaches can be
easily extended from system calls and network data to inputs from other
Intrusion Detection Systems or different sources of monitoring data. It could
also be easily adapted beyond HTC, for instance to monitor Cloud services
running in containers over orchestration engines such as Kubernetes and Mesos.

\par
We still have some limitations and topics for further analysis. We will
investigate optimizations that we can introduce in Arhuaco for reducing the
overhead. This seems feasible given the results for Linpack. Our framework has
not been tested in a production environment yet. The obtained False Positive
Rate, although small, is still significant if we consider the huge amount of
jobs running in the Worldwide LHC Computing Grid, close to 300.000 at any given
instant. We should improve the results in this area and also research into
methods to inform about the possible security incidents detected. We will
explore techniques to protect the privacy of our datasets in order to avoid
leaks of sensitive information. We will also investigate in how to employ the
distributed nature of the Grid to improve the distributed detection of
intrusions. This can be useful for instance, to autonomously inform other Grid
sites IDS about security incidents that could spread in the Grid. We are working
on further enhancing the isolation provided in our containers by using kernel
hardening such as grsecurity. Besides we are currently testing the integration
with other container solutions and HTC engines.

\section{Conclusions}
\label{conclusions}
We have presented Arhuaco, a security monitoring tool for High-Throughput
Computing. It employs Security by Isolation for executing and monitoring
Grid jobs inside Linux Containers based on multiple sources of monitoring data
such as network connections and system calls. It applies Convolutional Neural
Networks to classify Grid jobs data as normal or malicious. It also utilizes a
Recurrent Neural Network to learn a data model in order to generate new
training samples. The proposed algorithms implemented in Arhuaco improve the
security incident detection in Grid computing systems, since it is able to
identify the source of an intrusion with higher accuracy. Arhuaco can analyze
the huge flow of real time data that monitoring Grid jobs generate. This makes
Arhuaco suitable for environments such as the WLCG, the global Grid that
analyzes data from the Large Hadron Collider (LHC) at CERN, where millions of
jobs are executed every day. By a set of tests carried out in the ALICE Grid
as part of the WLGC, we have shown how the proposed algorithms outperform other
methods used in Intrusion Detection Systems for Grid Computing.

\section*{Datasets and software}
Arhuaco is under development. We are discussing the possibility to release it
and the training dataset as Open Source software and data.

\begin{acknowledgements}
Authors are grateful for the support provided by the ALICE offline team,
especially Costin Grigoras, Miguel Martinez Pedreira and the CERN security
department especially Stefan Lueders and Romain Wartel. Authors are also
grateful with the German Federal Ministry of Education and Research
(Bundesministerium fur Bildung und Forschung - BMBF) and HGS-HIRe for their
financial support.
\end{acknowledgements}

\bibliographystyle{spmpsci}       
\bibliography{biblio.bib}

\end{document}